\documentclass{article}
\usepackage[twocolumn]{emulateapj}
\submitted{To appear in the proceedings of the MPA/ESO 
conference ``The First Stars'' (August 4-7, 1999, Garching) ed.
A. Weiss etal. (Springer)}

\input epsf.sty

\newcommand\etal{{et al. }}
\newcommand\ms{$M_\odot$ }
\newcommand\msp{$M_\odot$}
\def\lsim{\mathrel{\rlap{\lower 4pt \hbox{\hskip 1pt $\sim$}}\raise 1pt \hbox
        {$<$}}}
\def\gsim{\mathrel{\rlap{\lower 4pt \hbox{\hskip 1pt $\sim$}}\raise 1pt \hbox
        {$>$}}}
\def\ee#1{$\times 10^{#1}$}
\def\ee2#1{\times 10^{#1}}

\lefthead{}
\righthead{Evolution and Nucleosynthesis of Metal-Free Massive Stars}

\begin{document}
\title{Evolution and Nucleosynthesis of Metal-Free Massive Stars }

\author{Hideyuki Umeda, Ken'ichi Nomoto, Takayoshi Nakamura}

\affil{ Department of
Astronomy, and Research Center for the Early Universe, 
School of Science, University of Tokyo, \\ Bunkyo-ku, 
Tokyo 113-0033, Japan \\E-mail: umeda@astron.s.u-tokyo.ac.jp; 
nomoto@astron.s.u-tokyo.ac.jp; nakamura@astron.s.u-tokyo.ac.jp}

\begin{abstract}

 We calculate presupernova evolutions and supernova explosions of
massive stars ($M=13-25 M_\odot$) for various metallicities.  We find
the following characteristic abundance patterns of nucleosynthesis in
the metal-free (Pop III) stars.  (1) The $\alpha$-nuclei (from C to
Zn) are more efficiently produced than other isotopes, and the
abundance pattern of $\alpha$-nuclei can be similar to the solar
abundance.  In particular, near solar ratios of alpha elements/Fe
might be a signature of Pop III which could produce a large amount of
Fe.  (2) The abundance ratios of odd $Z$ to even $Z$ elements such as
Na/Mg and Al/Mg become smaller for lower metallicity. However, these
ratios almost saturate below $Z \lsim 10^{-5}$, and [Na, Al/Mg] $\sim -
1$ for Pop III and low metal Pop II nucleosynthesis. This result is
consistent with abundance pattern of metal poor stars, in which these
ratios also saturate around $-1$. We suggest that these stars with the
lowest [Na/Mg] or [Al/Mg] may contain the abundance pattern of Pop III
nucleosynthesis.  (3) Metal poor stars show interesting trends in the
ratios of [Cr, Mn, Co/Fe]. We discuss that these trends are not
explained by the differences in metallicity, but by the relative
thickness between the complete and the incomplete Si burning
layers. Large [Co/Fe] and small [Cr, Mn/Fe] values found in the
observations are explained if mass cut is deep or if matter is ejected
 from complete Si burning layer in a form of a jet or bullets.  (4) We
also find that primary $^{14}$N production occurs in the massive Pop
III stars, because these stars have radiative H-rich envelopes so that
the convective layer in the He-shell burning region can reach the
H-rich region.

\end{abstract}

\section{Introduction}

Very metal poor stars provide important clues to investigate early
cosmic and galactic chemical evolution because their abundance
patterns would tell us what kind of stars (including the First Stars)
contribute to form those patterns.  We might be able to identify the
signature of the First Stars and thus can learn how the universe and
galaxies evolved chemically in its early phase.  Such identification
would be possible in the abundance pattern of an individual metal-poor
star, because in early phase of the galaxy evolution the ejecta of
each supernova may not have enough time to be uniformly mixed in the
galaxy (Audouze \& Silk 1995; Ryan \etal 1996; Shigeyama \& Tsujimoto
1998).

One example is the interesting trends of [Cr/Fe], [Mn/Fe] and [Co/Fe]
with [Fe/H] (McWilliam \etal 1995ab; Ryan \etal 1996; McWilliam 1997).
Since the published SNe II yields (Woosley \& Weaver 1995, hereafter
WW95; Thielemann, Nomoto \& Hashimoto 1996; Hashimoto 1995; Tsujimoto
\etal 1995; Nomoto \etal 1997) 
are not consistent with these trends, one might suspect
that these trends would be a signature of Pop III star yields.

Here we define the First Stars as metal-free (Pop III) stars.  We
evolve and explode massive metal-free stars as well as other
metallicity stars including the metallicity dependent mass loss.  We
calculate nucleosynthesis to see what are the indications of the
ejecta of the first stars.  Previously, WW95, Arnett (1996), and
Limongi \etal (1999, LCS) calculated nucleosynthesis of SNe II using
for various metallicities down to metal-free, but these studies did
not include mass loss and did not examine the dependence on the
convective mixing.

In addition to the metallicity effects, we study the dependence on the
explosion energy.  This is because some recent supernovae have been
recognized to be hyper-energetic (hypernovae) being more than $10^{52}$
ergs.  The examples include SN 1998bw (Galama \etal 1998; Iwamoto
\etal 1998; Woosley \etal 1999), SN 1997ef (Iwamoto \etal 1999), and
SN 1997cy (Germany \etal 1999; Turatto \etal 1999).  These hypernovae
are suggested to originate from quite massive stars, so that their
nucleosynthesis may significantly contribute to the early cosmic and
galactic chemical evolution.

Our results are also sensitive to the efficiency of convective mixing.
Here we report mostly the case of low efficiency, and more complete
results will be presented in a forthcoming paper.

\section{Code and Method}

 We have developed a stellar evolution code which is based on Henyey
type stellar evolution codes by Nomoto \& Hashimoto (1988, NH88), Saio
\& Nomoto (2000), and Umeda \etal (1999).  We start calculations from
the zero-age main-sequence through core collapse including metallicity
dependent mass loss (de Jager \etal 1988; Kudritzki \etal 1989).  For
example, the star of initially 25 \ms with solar metallicity 
decreases its mass down to 21.5 \ms before the explosion.

We adopt the OPAL opacity as a function of metallicity, $X$(H), $X$(C)
and $X$(O) (Iglesius \& Rodgers 1993).  This code runs a nuclear
reaction network by Hix \& Thielemann (1996) for the calculation of
nuclear energy generation and nucleosynthesis. For nucleosynthesis
of H and He burning, 51 isotopes up to Si are included, and after He
burning 240 isotopes up to Ge are included.

Schwartzshild criterion is adopted for convective stability criterion
and diffusive convective mixing by Spruit (1992) is adopted. In this
work, a case for relatively slow mixing, $f_k=0.05$ (Saio \& Nomoto
2000), is investigated.

 At the end of core helium burning, the central C/O ratio is largely
influenced by the uncertain ${\rm ^{12}C(\alpha,\gamma)^{16}O}$
reaction rate (Fowler 1984).  The C/O ratio affects the abundances of
Ne, Mg and Al, and also the Fe core mass.

In this work, the $^{12}$C$(\alpha,\gamma)^{16}$O reaction rate is
chosen to be 1.4 times the value given in Caughlan \& Fowler (1988;
CF88).  With this choice for the 25 \ms model of $Z$=0.02, the central
C/O ratio is 0.2 at the end of core helium burning, which is close to
the NH88 models.

 Using the presupernova models, we carry out 1D hydrodynamical
simulations of SN II explosions with a PPM code which uses a small
nuclear reaction network.  Then the explosive nucleosynthesis is
calculated as a post-processing using a larger reaction network of 300
isotopes up to Br as described in Thielemann \etal (1996).

\section{Presupernova Stellar Evolution} 

\begin{figure}
\hskip 3 cm 
\epsfxsize=12.cm
\epsfysize=10.cm
\epsfbox{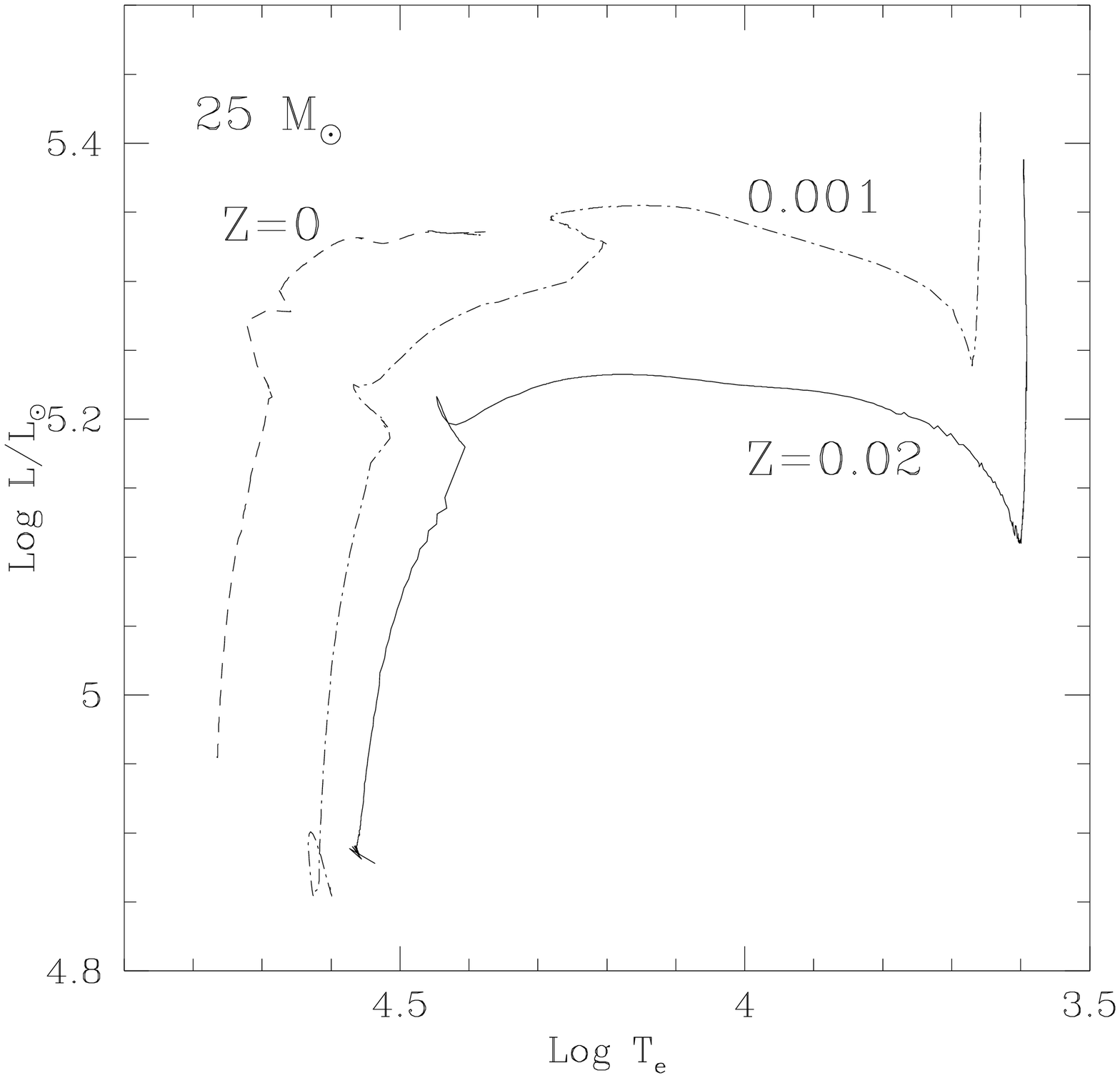}
\figcaption{The H-R diagram of the 25 \ms stars with various metallicities.
\label{HR25}}
\end{figure}

 Figure \ref{HR25} shows the H-R diagram of the 25 \ms star with
metallicity of $Z=0, 0.001$, and 0.02. As shown in this figure, the
star is more luminous for lower metallicity owing to the smaller
opacity. The evolution in the H-R diagram of zero metal (Pop III)
stars with 13 \ms $\lsim M \lsim$ 25 \ms are quite different from Pop
I and II stars, because they do not evolve to red-giants before core
collapse (e.g., Castellani \etal 1983).

\begin{figure}
\hskip 3cm
\epsfxsize=12cm
\epsfysize=8cm
\epsfbox{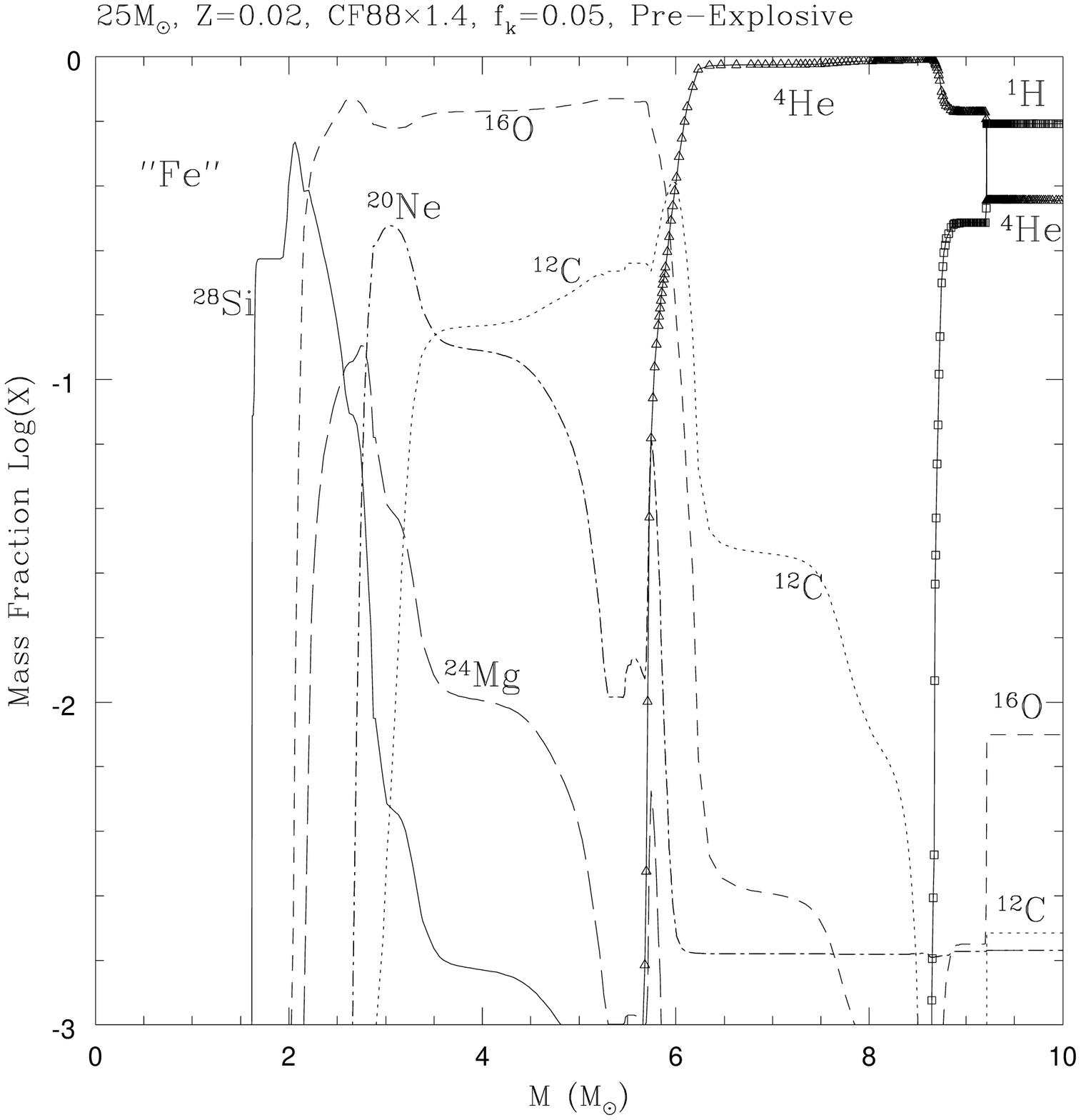}

\hskip 3cm
\epsfxsize=12.cm
\epsfysize=8cm
\epsfbox{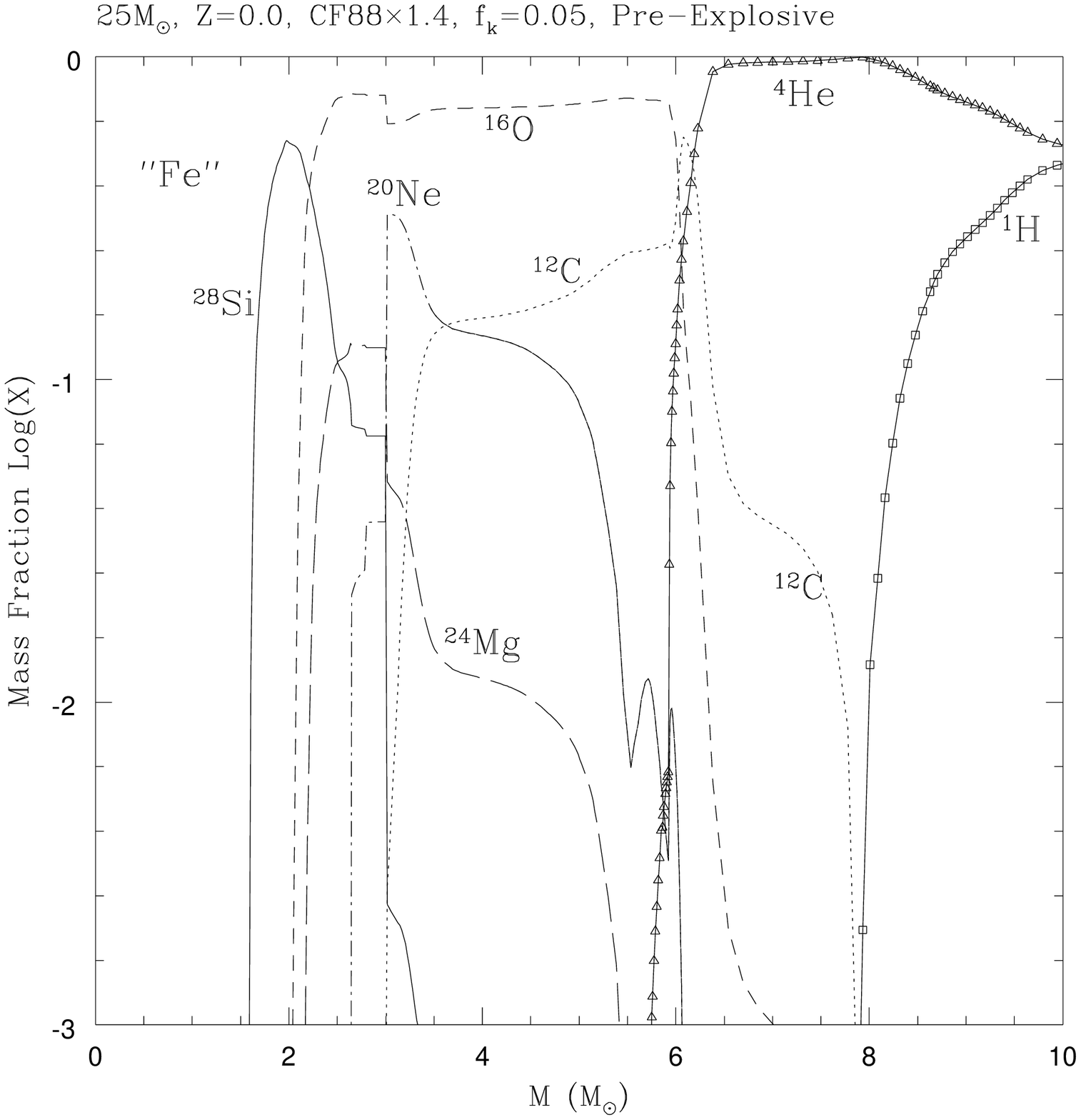}

\hskip 3cm
\epsfxsize=12cm
\epsfysize=8cm
\epsfbox{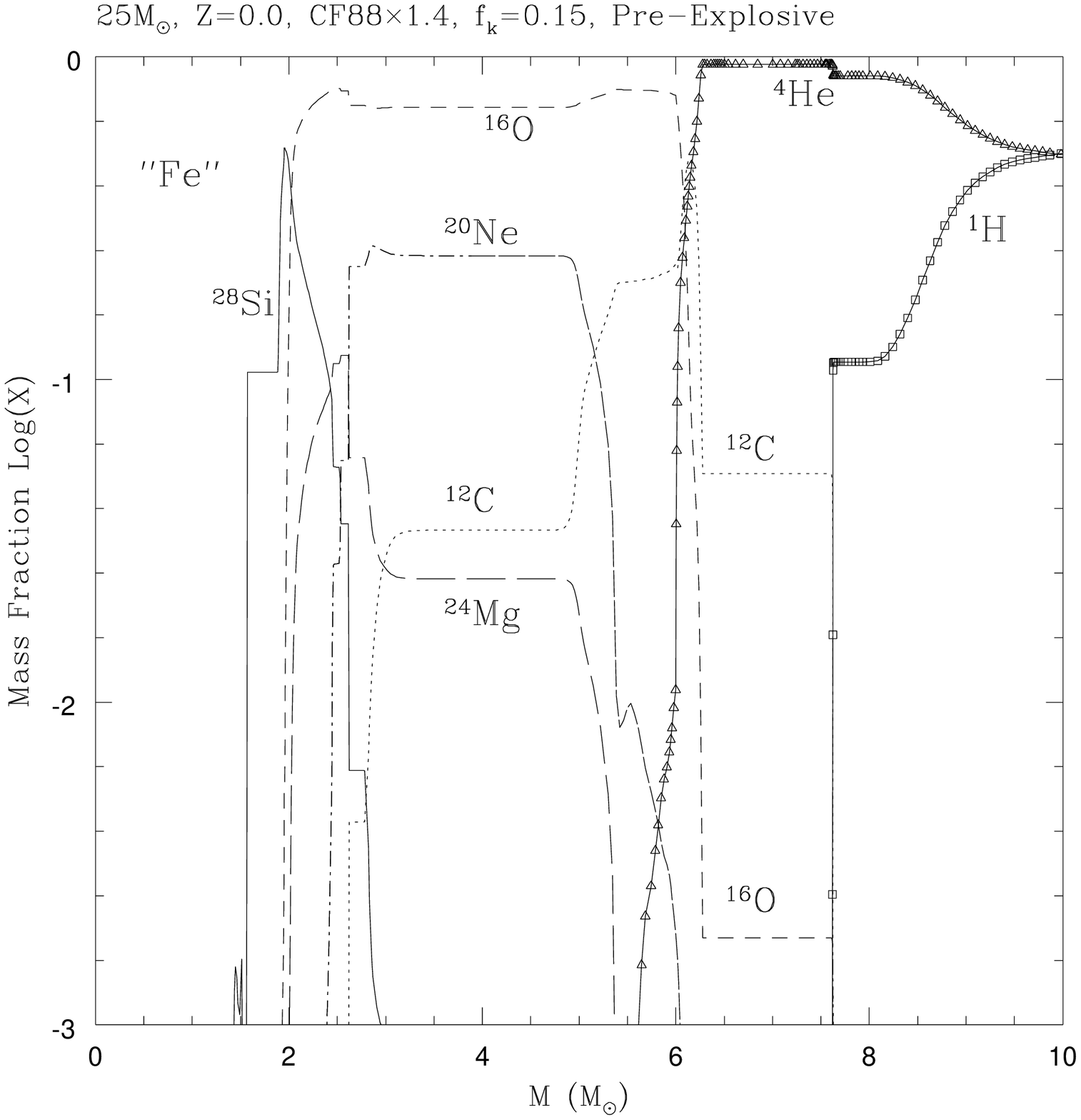}

\caption{Presupernova abundance distribution. The top and the middle
panels show the cases with $Z=0.02$ and $Z=0$, respectively.  The
bottom one is the $Z=0$ case with a more efficient convective mixing
($f_k=0.15$) than others. \label{pre25}}
\end{figure}

Figures \ref{pre25} shows abundance distribution of presupernova
stars.  The H-rich envelope for $Z=0.02$ is convective, while the
H-rich envelope for $Z=0$ is radiative.  For $Z=0$, only a small
amount of CNO elements are produced in the H-rich layer, so that
hydrogen burning is too weak to drive the expansion toward a
red-giant.

  For Pop I and II stars, lower metallicity leads to the formation of
a more massive core because of the larger luminosity. One might expect
Pop III stars have the largest cores because of this effect; however,
the He core size is smaller than Pop II because of weak H-shell
burning.  In our model these two effects cancel out and the masses of
the He and C-O cores are almost the same between $Z=0$ and $Z=0.02$
models (Fig. \ref{pre25}).


\subsection{Fe Core Mass} 

 The Fe core mass is mostly determined by the He (or C-O) core mass,
and the C/O ratio after the central helium burning. In general, a
larger He (or C-O) core mass leads to a larger Fe core mass. The
larger C/O ratio leads to weaker central O-burning and a smaller Fe
core mass.

A more massive or lower (but non-zero) metallicity star tends to have
a larger luminosity and hence a larger He (and C-O) core mass. Also
the C/O ratio is smaller for a larger core mass. Therefore, the Fe
core mass tends to be larger for a larger stellar mass and smaller
metallicity, although its size also depends on the behavior of
convection and these relations do not hold always.

\begin{figure}
\hskip 3cm
\epsfxsize=12.cm
\epsfysize=7.cm
\epsfbox{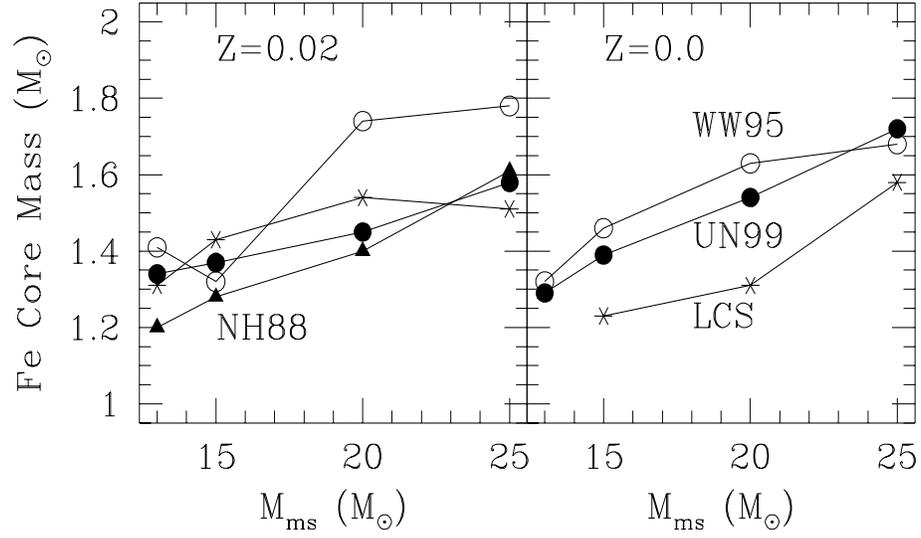}

\caption{The Fe core mass as compared with other works.
\label{fecore1}}
\end{figure}

 Figure \ref{fecore1} summarizes the masses of the presupernova Fe
cores compared with other works. We should note that all other works
did not include mass loss.


 Although opacity is smaller, Pop III stars have smaller core masses
than Pop II stars for the same main-sequence mass because H-burning
via the CNO cycle is weaker. Since the Pop II stars have larger cores
than Pop I, the core masses of Pop III are closer to Pop I rather than
Pop II. Note that this relation does not always hold since the core
mass depends on the behavior of convection which is rather chaotic
(Barkat \etal 1990).

\subsection{$Y_{\rm e}$ distribution} 

\begin{figure}
\hskip 3cm
\epsfxsize=12.cm
\epsfysize=10.cm
\epsfbox{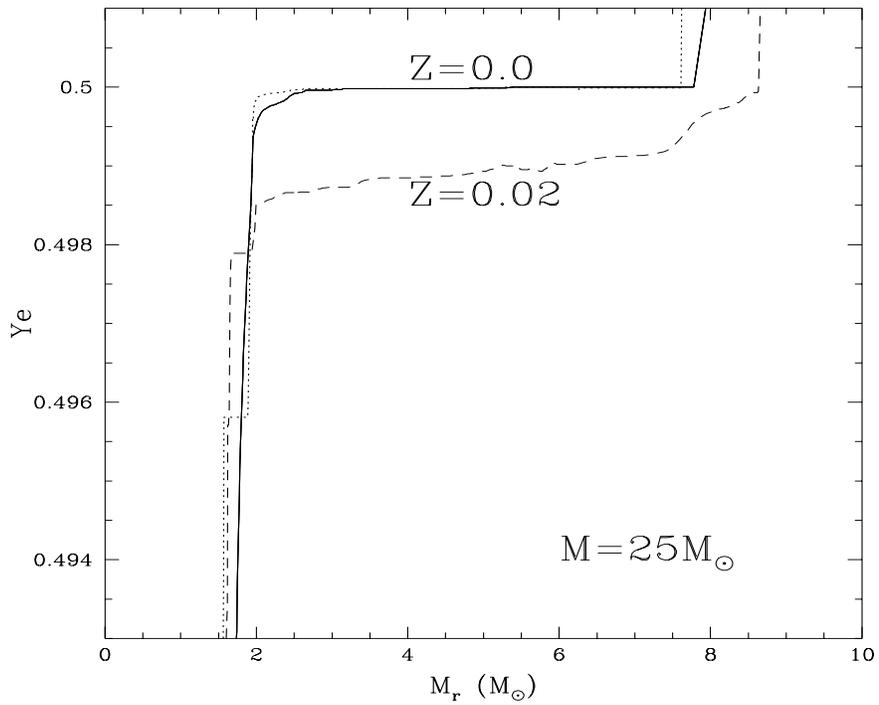}

\caption{$Y_e$ distribution of the pre-explosive 25 \ms stars for $Z=0$
and 0.02.  The solid and dotted lines corresponds to $f_k=0.05$ and 0.15
models, respectively.
\label{ye1}}
\end{figure}

 The electron mole number $Y_e$ is important for the explosive
nucleosynthesis.  In Figure \ref{ye1} we show the distribution of
$Y_e$ in the presupernova model.  This shows that $Y_e$ is larger for
zero metal stars than solar metallicity stars, because of smaller
$^{22}$Ne which is produced from $^{14}$N.

\section{Explosive nucleosynthesis}

     The general features of the explosive nucleosynthesis in SNe II
are summarized as follows (e.g., Woosley \etal 1973; Hashimoto \etal
1989; Thielemann et al. 1994, 1996).  The explosively produced
elements depend mainly on the peak temperature attained through the
passage of the shock.  The region after the shock passage is radiation
dominant, so that the peak temperature is approximately related to the
stellar radius $r$ and the deposited energy $E^*$ as

\begin{equation}
  T_9 = (E^*_{51})^{1/4} \ (r/3.16\ee2 4 \ {\rm km})^{-3/4},
\end{equation} 
\noindent 
where $E^*_{51}$ is the deposited energy in units of $10^{51}$ erg,
$T_9$ is the peak temperature in $10^9$ K.  Therefore, assuming
$E^*_{51}=1$, explosive nuclear burning are classified into several
cases according to $T_9$ (or $r$) at $T_9=5$ ($r$ = 3700 km), $T_9=4$
($r$ = 4980 km), $T_9=3.3$ ($r$ = 6430 km), and $T_9=2.1$ ($r$ =
11,800 km).  The relation between the presupernova radius $r$ and the
enclosed mass $M_r$ depends on the progenitor's mass and metallicity
(e.g., Nomoto \etal 1993).

\subsection{Complete Silicon Burning and Neutron-rich Species}

     For $T_{\rm 9} >$ 5, explosive silicon and oxygen burning leads
to nuclear statistical equilibrium and produces mostly $^{56}$Ni if
$Y_{\rm e} > 0.49 $, $T_9 < 8$, and $\rho < 10^8 $ g cm$^{-3}$.  Other
iron peak elements are also produced owing to this high temperature but
the final products depend on the time scale of the freezeout and
initial distribution of $Y_{\rm e}$.

     For the inner region, $\alpha$-rich freeze out occurs after
complete silicon burning.  Some ${\rm ^4He}$ remain even after the
freezeout without recombining into the iron peak elements.  In these
regions, appreciable amount of radioactive nuclei, ${\rm ^{56}Ni,
^{57}Ni, and ^{44}Ti}$, are produced.  Also ${\rm ^{56,57}Fe}$ and ${\rm
^{58,60,62}Ni}$ are synthesized here.  ${\rm ^{60,62}Ni}$ are originally
produced as ${\rm ^{60,62}Zn}$ which decay as ${\rm ^{60,62}Zn
\rightarrow ^{60,62}Cu \rightarrow ^{60,62}Ni}$.  These neutron-rich
nuclei ($Y_{\rm e}$ = 0.484 for ${\rm ^{62}Zn}$ and $Y_{\rm e}$ = 0.483
for $^{58}$Ni) are produced in the neutron-rich layers near the mass cut
in SNe II as well as in Type Ia supernovae.

     In smaller mass stars, a larger amount of neutron-rich elements
$^{58}$Ni, $^{60,62}$Zn are produced near the mass cut.  The mass cut
is related to the amount of radioactive nuclei.  The most important
ones are $^{56}$Ni and ${\rm ^{57}Co}$ (${\rm ^{57}Ni \rightarrow
^{57}Co \rightarrow ^{57}Fe}$) as actually observed in SN 1987A (e.g.,
Arnett \etal 1989; Nomoto \etal 1994) and the obtained ratio
$^{56}$Ni/$^{57}$Ni was used to infer the mass cut (Kumagai \etal
1989, 1993).

\smallskip

\subsection {Incomplete Silicon Burning and Explosive Oxygen Burning}

     For 4 $< T_{\rm 9} <$ 5, incomplete silicon burning and
explosive oxygen burning produce mostly ${\rm ^{28}Si, ^{32}S,
^{36}Ar}$ and ${\rm ^{40}Ca}$.  Some $^{44}$Ti and $^{56}$Ni are also
produced by incomplete silicon burning.  Oxygen is burned up
completely and $^{28}$Si, $^{32}$S and other $\alpha$-particle nuclei
are produced.  The peak temperature is too low to produce iron peak
elements.

     For 3.3 $< T_{\rm 9} <$ 4.0, explosive neon burning produces
some amounts of ${\rm ^{16}O,^{28}Si}$, and ${\rm ^{32}S}$.

     For 2.0 $< T_{\rm 9} <$ 3.3, explosive carbon burning produces
${\rm ^{20}Ne, ^{24}Mg}$.  However, the peak temperature is too low to
change the initial abundances appreciably.  For $T_{\rm 9} <$ 2, no
significant explosive burning occurs.  For the oxygen-rich layer with
2 $< T_{\rm 9} <$ 3, where explosive neon and carbon burning take
place, the p--process occurs via the photodisintegration of nuclei:
($\gamma$, p), ($\gamma$, n), ($\gamma, \alpha)$ (e.g., Rayet \etal
1995).

     The inner part of the star where $T_{\rm 9} >$ 2 undergoes the
explosive nucleosynthesis according to the above classification.  The
outer part, including most of the original oxygen-rich layer, is
ejected without being processed by explosive nuclear burning.  The
ratio between the explosive and hydrostatic burning products is
sensitive to the stellar mass.

\section {Abundance Distribution}

\begin{figure}
\hskip 3cm
\epsfxsize=12.cm
\epsfysize=11.cm
\epsfbox{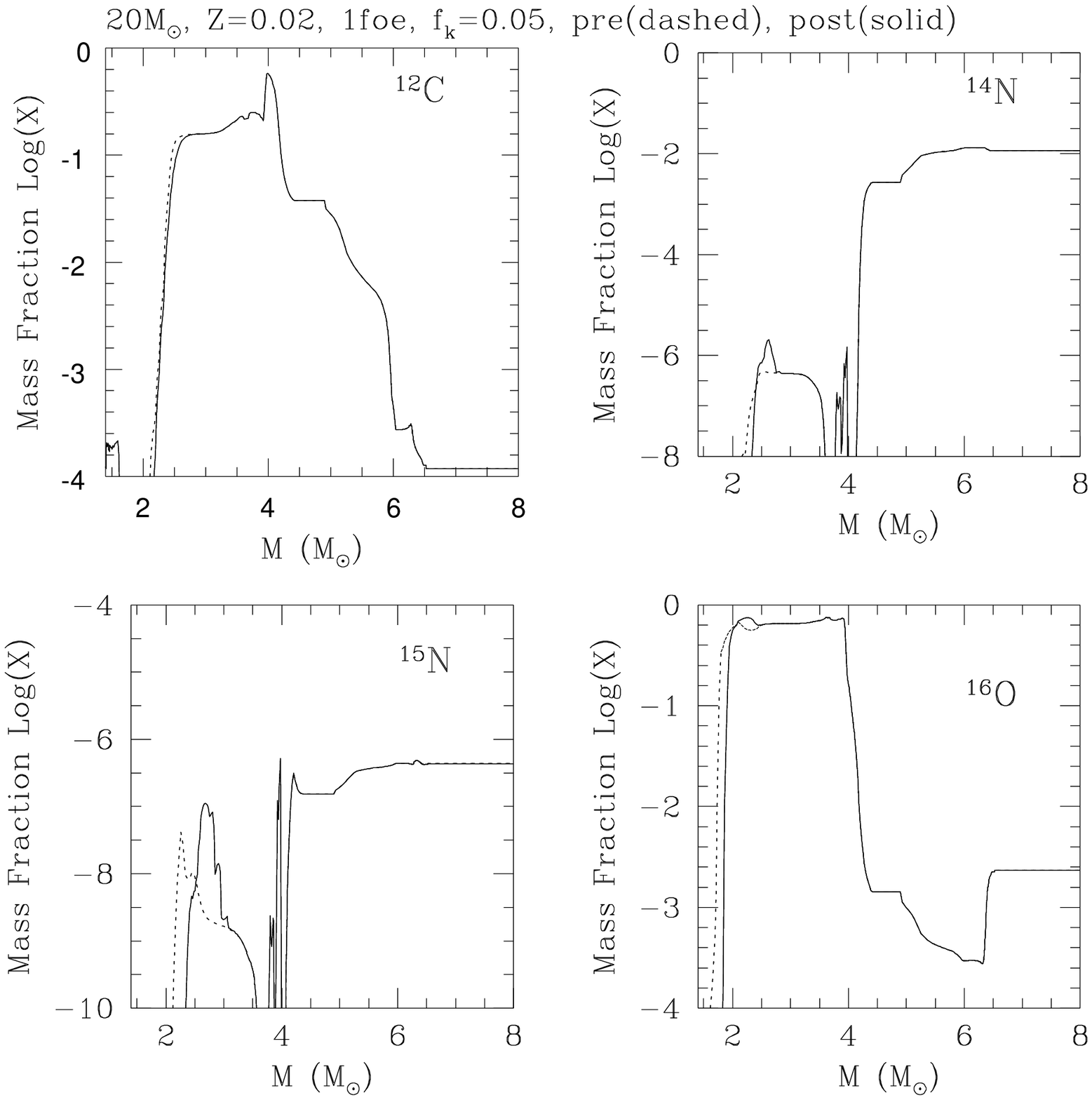}

\hskip 3cm
\epsfxsize=12.cm
\epsfysize=11.cm
\epsfbox{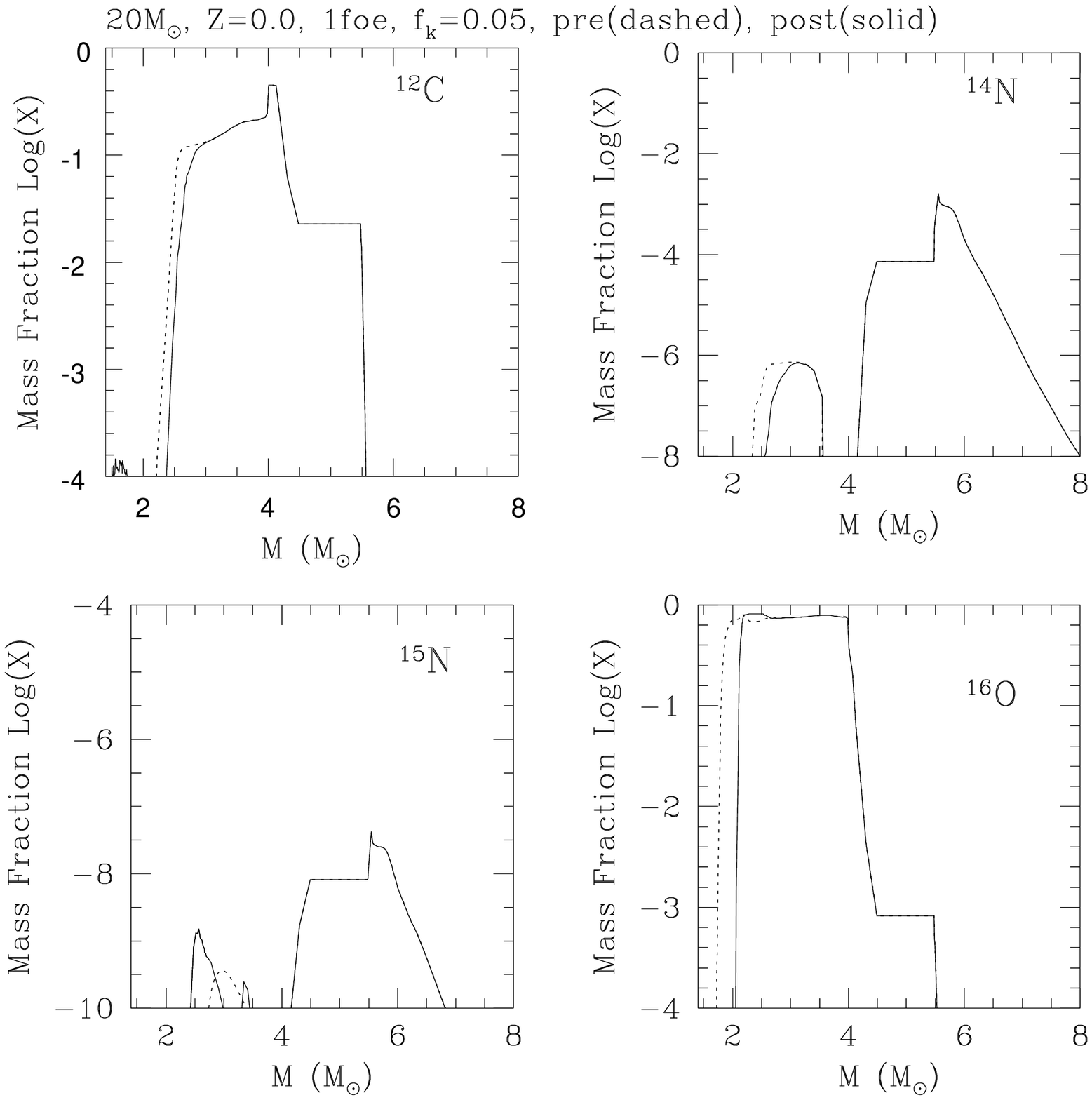}
\caption{Pre and post-explosive abundance distribution 
of $^{12}$C, $^{14}$N, $^{15}$N, $^{16}$O
for 20\msp, $Z$=0.02 (upper 4 panels)
and $Z$=0 (lower 4 panels) models.\label{dist1}}
\end{figure}

\begin{figure}
\hskip 3cm
\epsfxsize=12.cm
\epsfysize=11.cm
\epsfbox{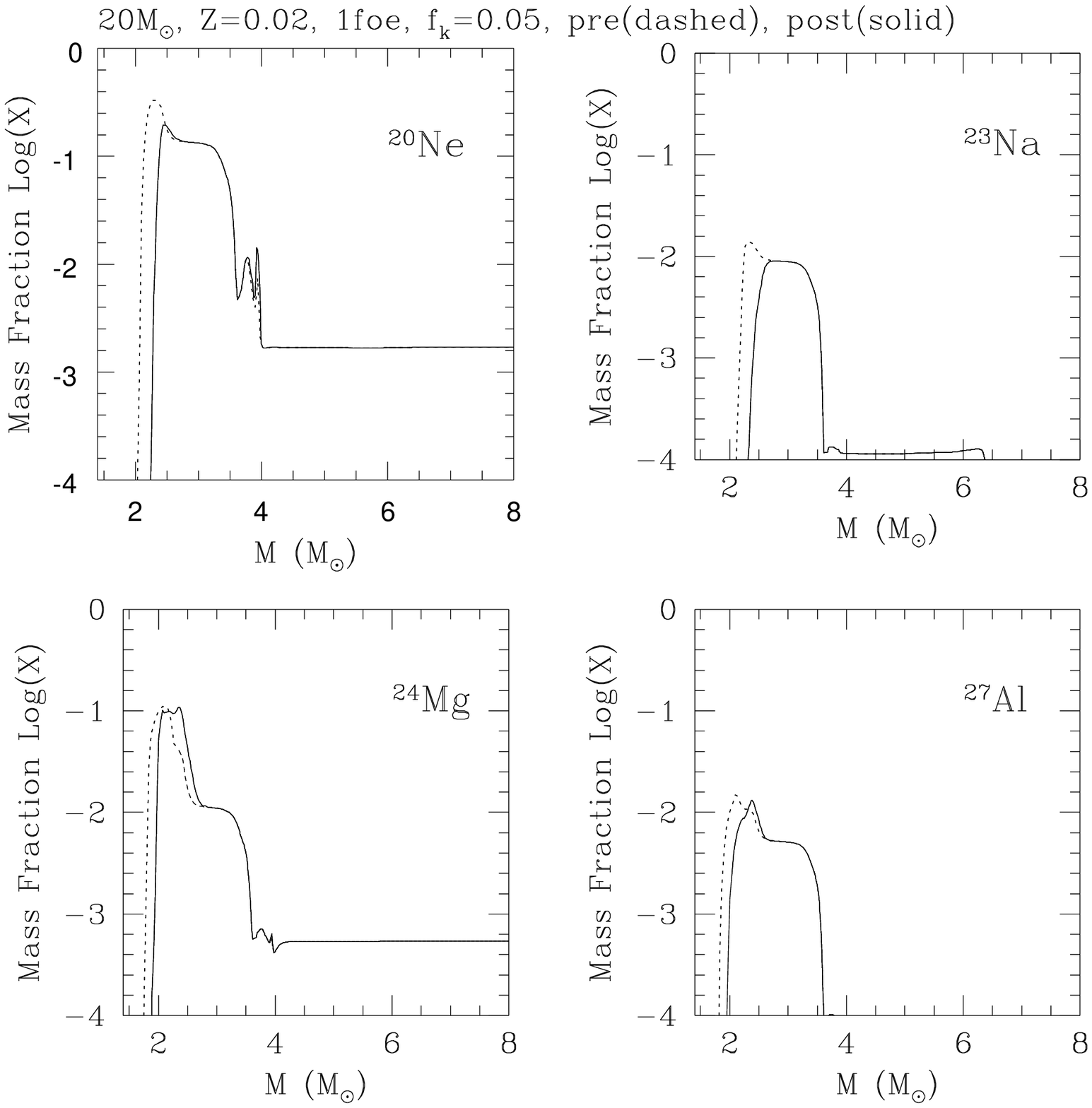}

\hskip 3cm
\epsfxsize=12.cm
\epsfysize=11.cm
\epsfbox{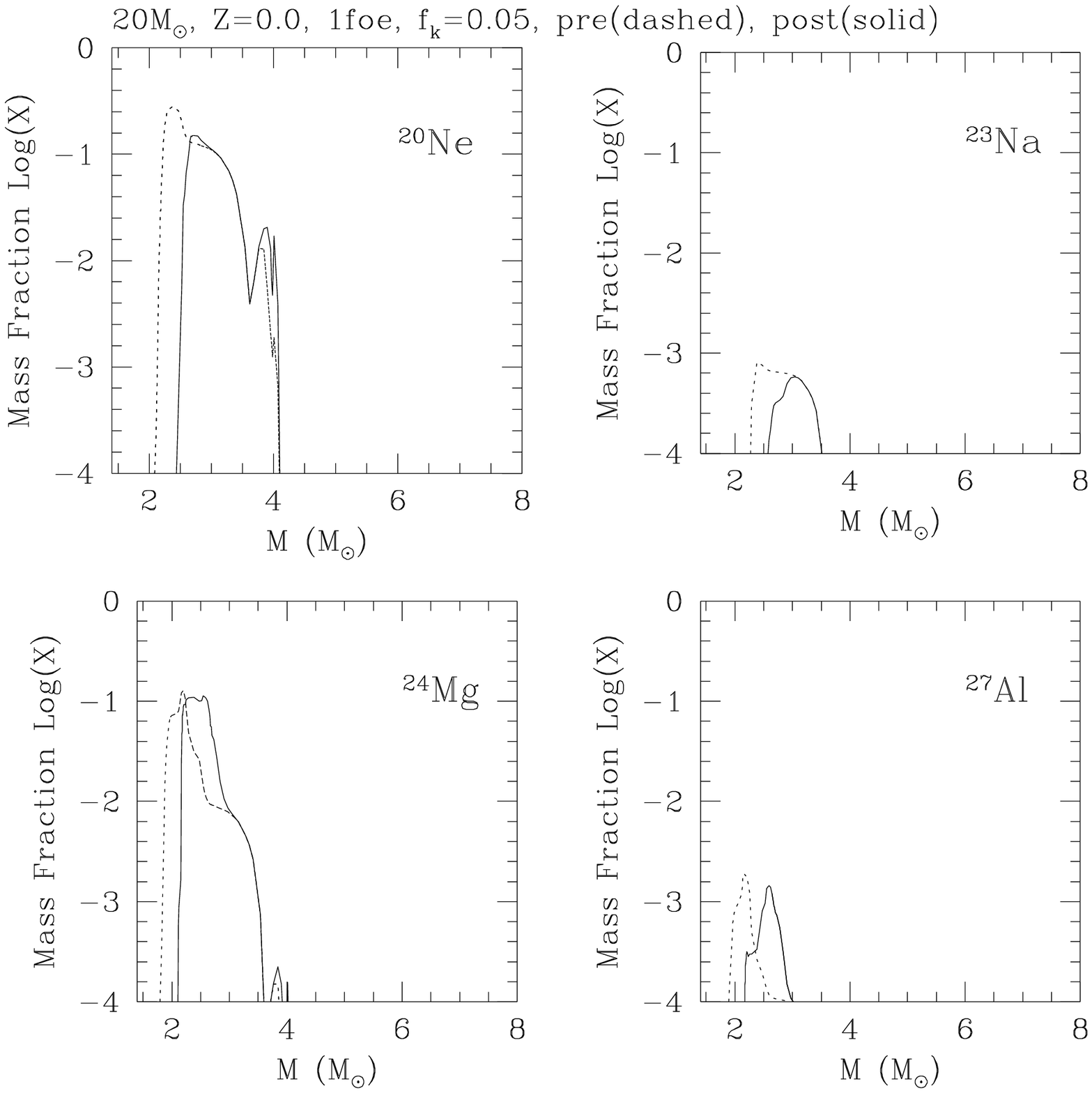}
\caption{Pre and post-explosive abundance distribution 
of $^{20}$Ne, $^{23}$Na, $^{24}$Mg, $^{27}$Al
for the 20 \msp, $Z=0.02$ (upper 4 panels)
and $Z=0$ (lower 4 panels) models.\label{dist2}}
\end{figure}

\begin{figure}
\hskip 3cm
\epsfxsize=12.cm
\epsfysize=11.cm
\epsfbox{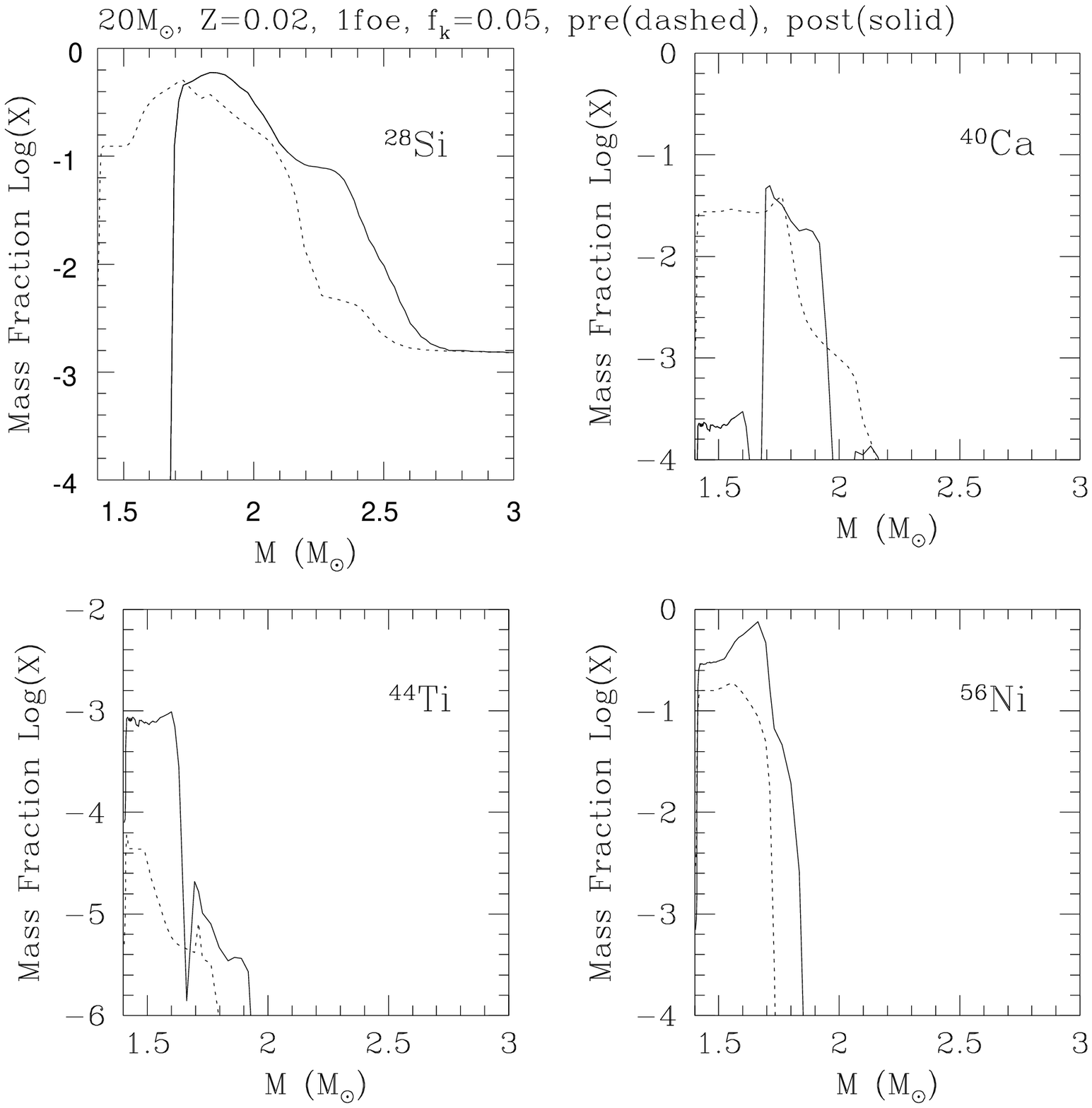}

\hskip 3cm
\epsfxsize=12.cm
\epsfysize=11.cm
\epsfbox{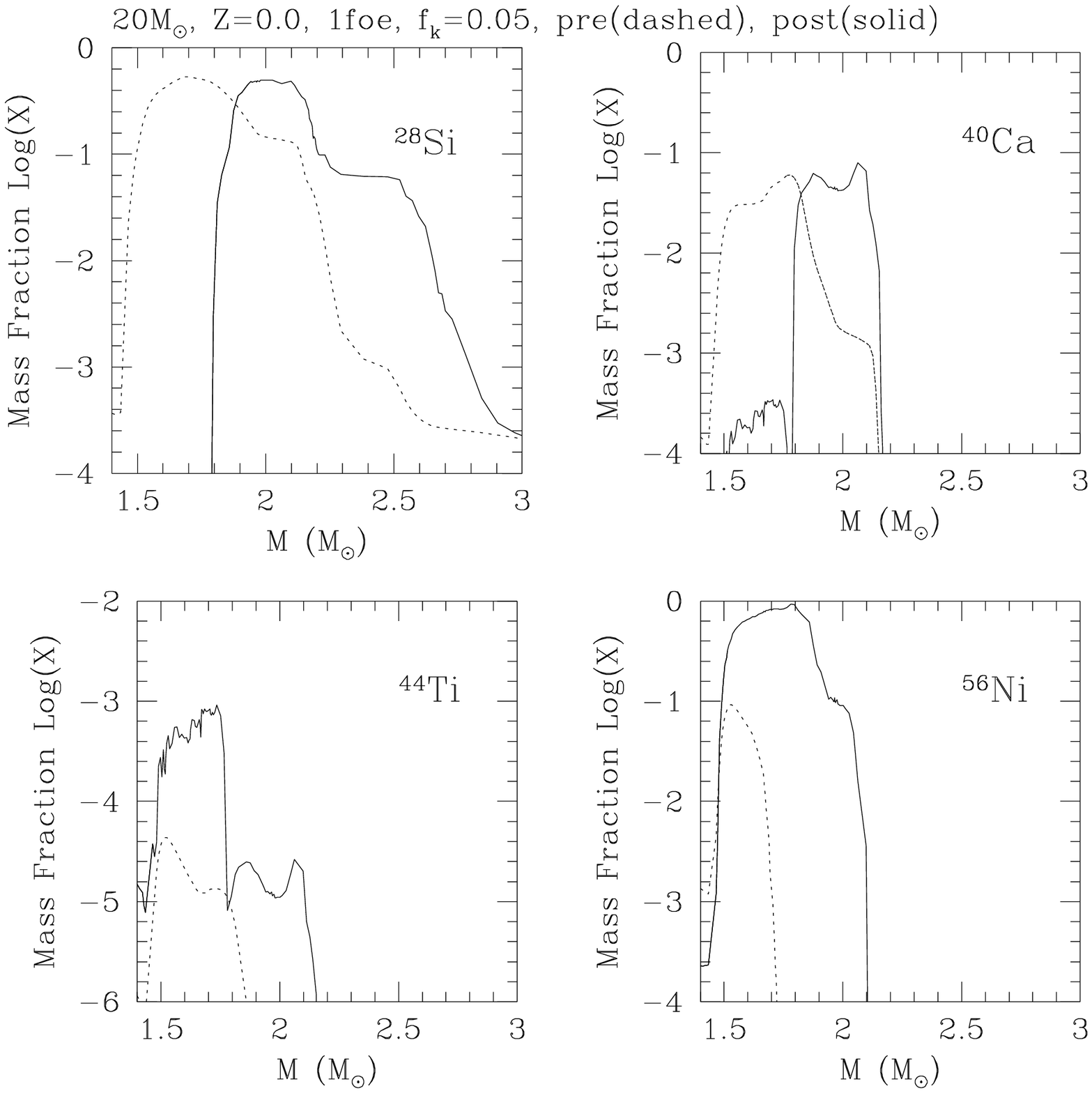}
\caption{Pre and post-explosive abundance distribution 
of $^{28}$Si, $^{40}$Ca, $^{44}$Ti, $^{56}$Ni
for the 20 \msp, $Z=0.02$ (upper 4 panels)
and $Z=0$ (lower 4 panels) models.\label{dist3}}

\end{figure}

\begin{figure}
\hskip 3cm
\epsfxsize=12.cm
\epsfysize=10.cm
\epsfbox{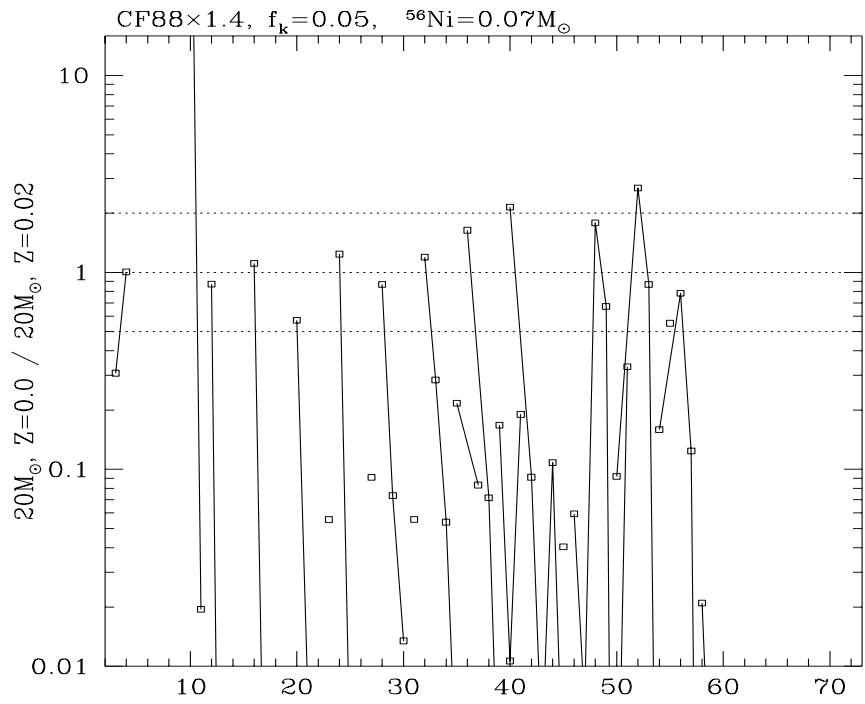}
\caption{Comparison of the 20 \ms star yields between $Z=$ 0 and 0.02.
\label{z0z002comp}}
\end{figure}

     Figures \ref{dist1}--\ref{dist4} show the abundance distribution
of some isotopes before and $10^3$ sec after explosive nuclear burning
for the 20 \ms star with $Z=0$ and 0.02. These figures show that $Z=0$
and 0.02 models yield roughly the same amount of alpha nuclei which
are produced during He and C-burnings such as $^{12}$C, $^{16}$O,
$^{20}$Ne, $^{24}$Mg, and produced during explosive burning such as
$^{40}$Ca, $^{44}$Ti, $^{52}$Fe, $^{64}$Ge.  On the other hand, other
elements produced during H, He and C-burnings such as $^{13}$C,
$^{14}$N, $^{15}$N, $^{23}$Na, $^{27}$Al, $^{31}$P are much less
abundant in the $Z=0$ model compared with the $Z=0.02$ model. This
also can be seen in Figure \ref{z0z002comp}. 

     As examples of these two types of species, we discuss the
ejected masses of radioactive species $^{26}$Al and $^{44}$Ti (decays
into $^{44}$Ca), which are important for the Line Gamma-Ray astronomy.
The $^{44}$Ti mass is in general roughly independent of metallicity,
but depends sensitively on the mass cut.  In Figure \ref{z0z002comp},
$^{44}$Ti and all heavy elements with $A\geq 60$ appear to be much
less abundant for $Z=0$ than $Z=0.02$.  This is only because the mass
cut is set to produce 0.07 \ms $^{56}$Ni for both cases despite the
difference in the $^{56}$Ni distribution; thus the abundances of those
isotopes produced by complete Si burning relative to $^{56}$Ni appear
to be smaller for $Z=0$ ($\sim 10^{-5}$ \msp) than $Z=0.02$ by a
factor of 2 - 4.  We stress again that the $^{44}$Ti mass is uncertain
because it significantly depends on mass cut.

     $^{26}$Al is produced in the H, He and C-burnings and the ejected
mass is independent of mass cut, while its mass is smaller for smaller
metallicity.  For $Z=0.02$, the $^{26}$Al mass ranges from 1 $\ee2{-5}$
\ms to 8 $\ee2{-5}$ \ms for the 13 to 25 \ms stars.  For $Z=0$, these
masses range from 1 $\ee2{-7}$ \ms to 2 $\ee2{-6}$ \msp.

\section {Integrated Abundances Relative to the Solar Abundances}

\begin{figure}
\hskip 3cm
\epsfxsize=12.cm
\epsfysize=11.cm
\epsfbox{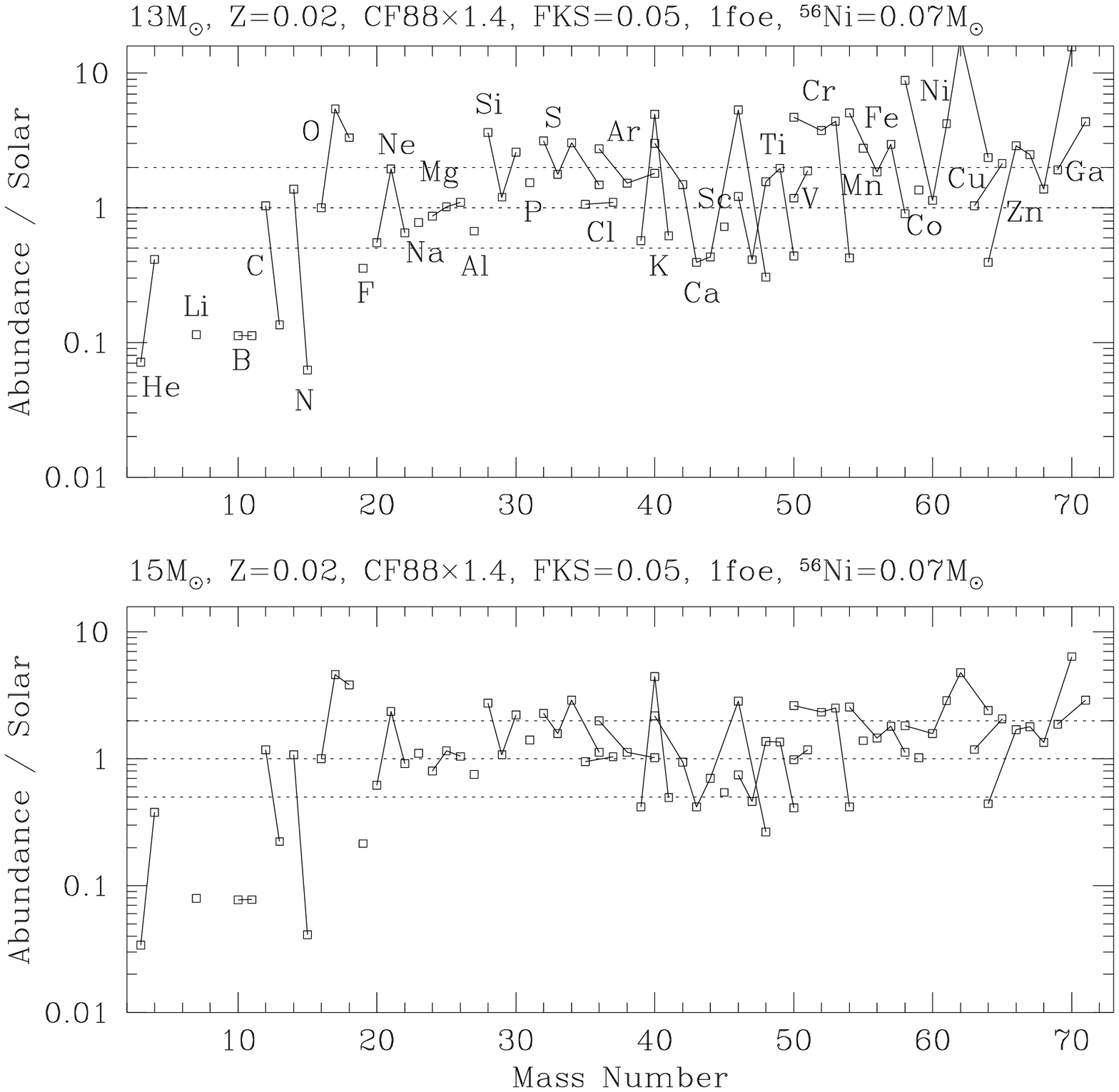}

\hskip 3cm
\epsfxsize=12.cm
\epsfysize=11.cm
\epsfbox{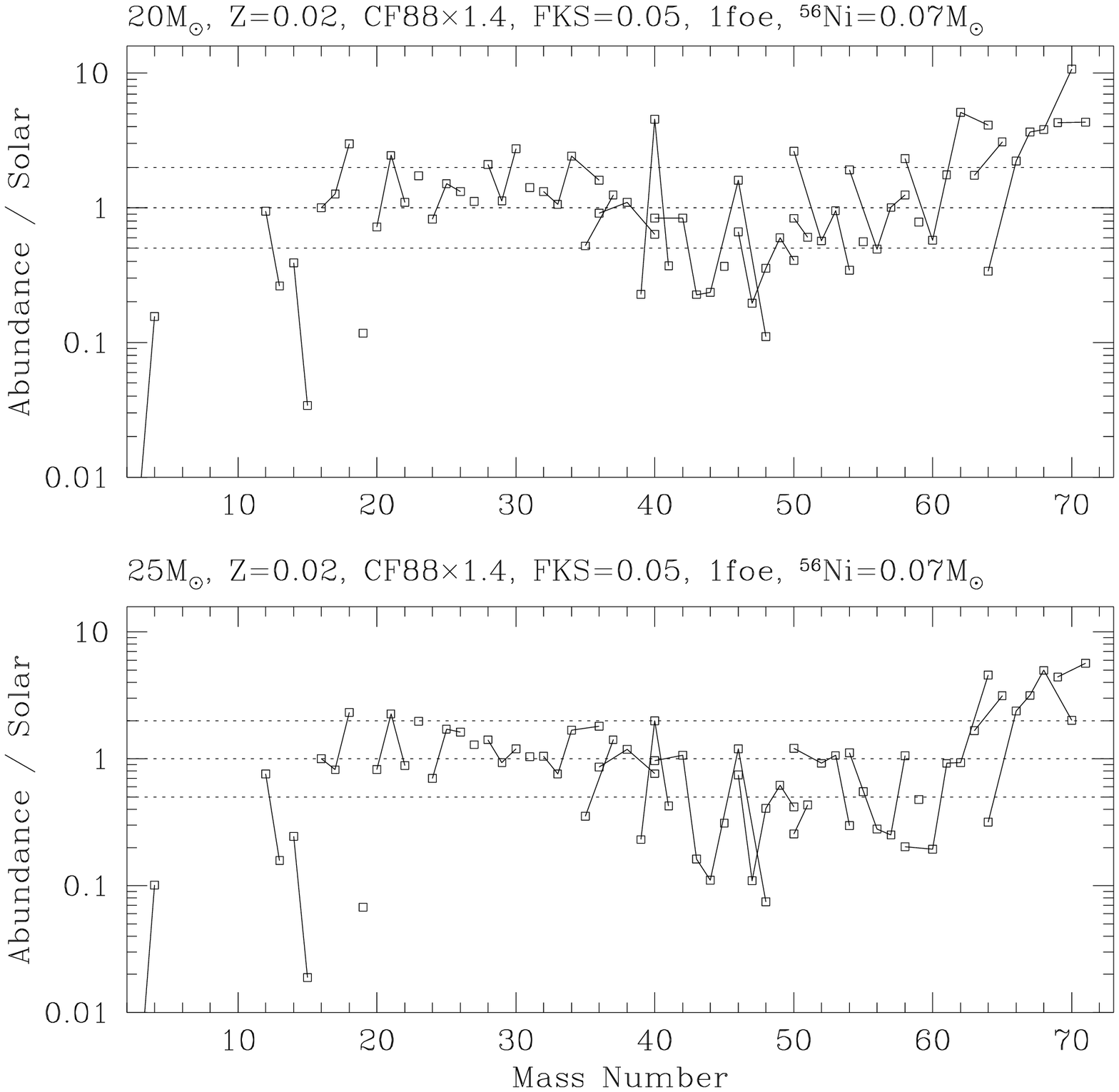}
\caption{Abundances for $Z=0.02$, $M=13-25$ \ms
models normalized by the solar abundances.
\label{abunz002}}
\end{figure}

\begin{figure}
\hskip 3cm
\epsfxsize=12.cm
\epsfysize=11.cm
\epsfbox{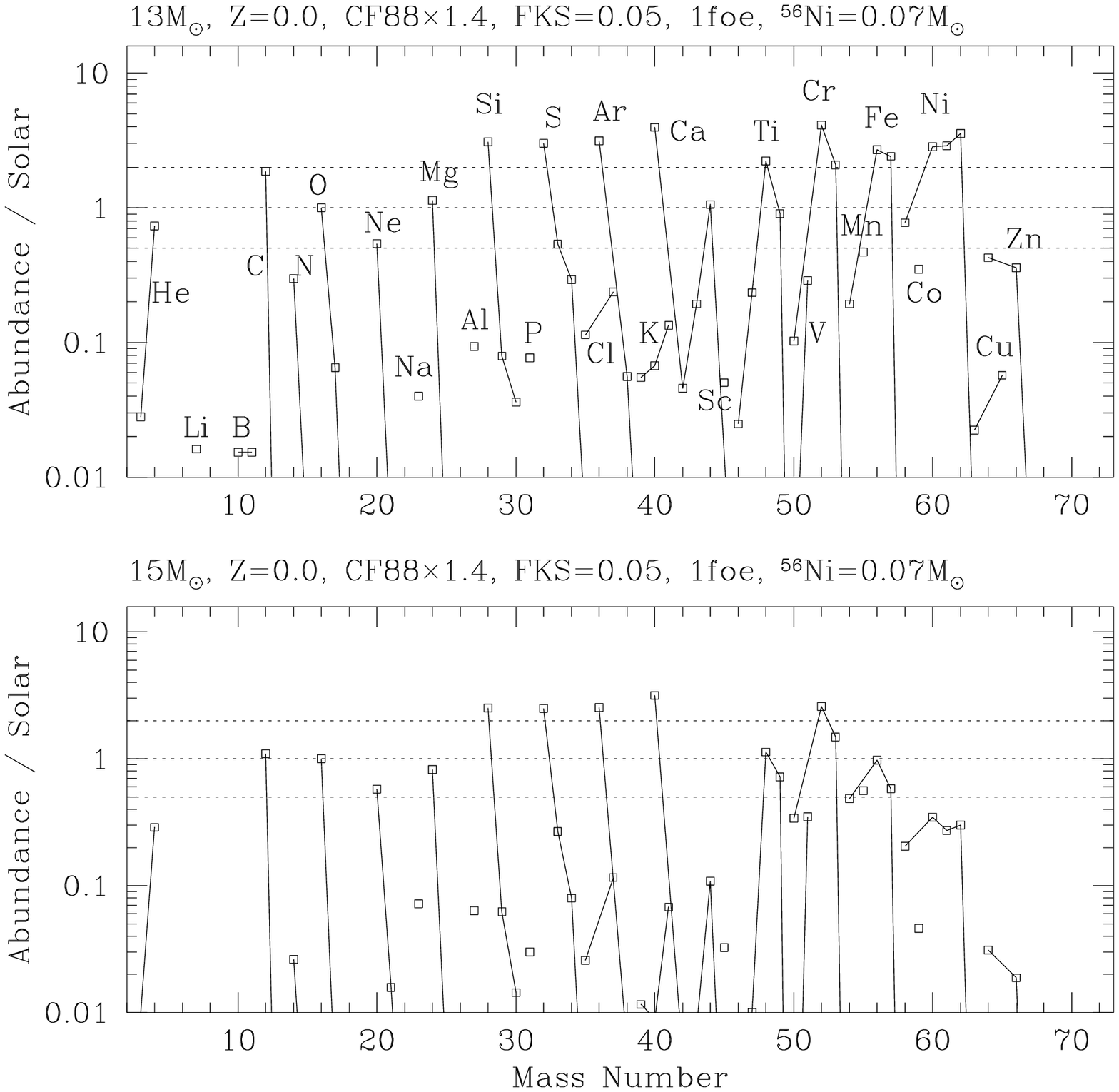}

\hskip 3cm
\epsfxsize=12.cm
\epsfysize=11.cm
\epsfbox{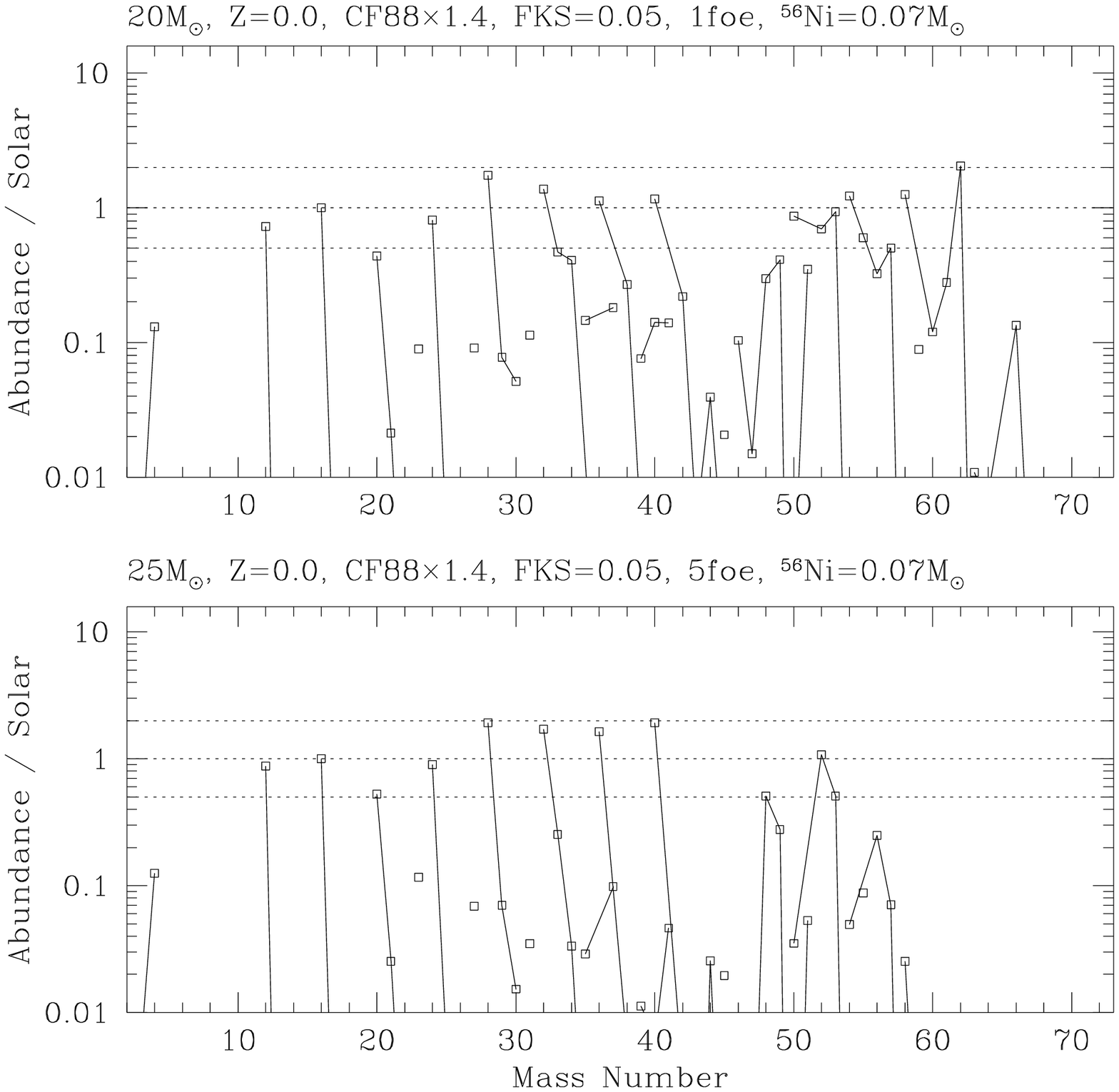}
\caption{Abundances for $Z=0$, $M=13-25$ \ms
models normalized by the solar abundances.
\label{abunz0}}
\end{figure}

\begin{figure}
\hskip 3cm
\epsfxsize=12.cm
\epsfysize=11.cm
\epsfbox{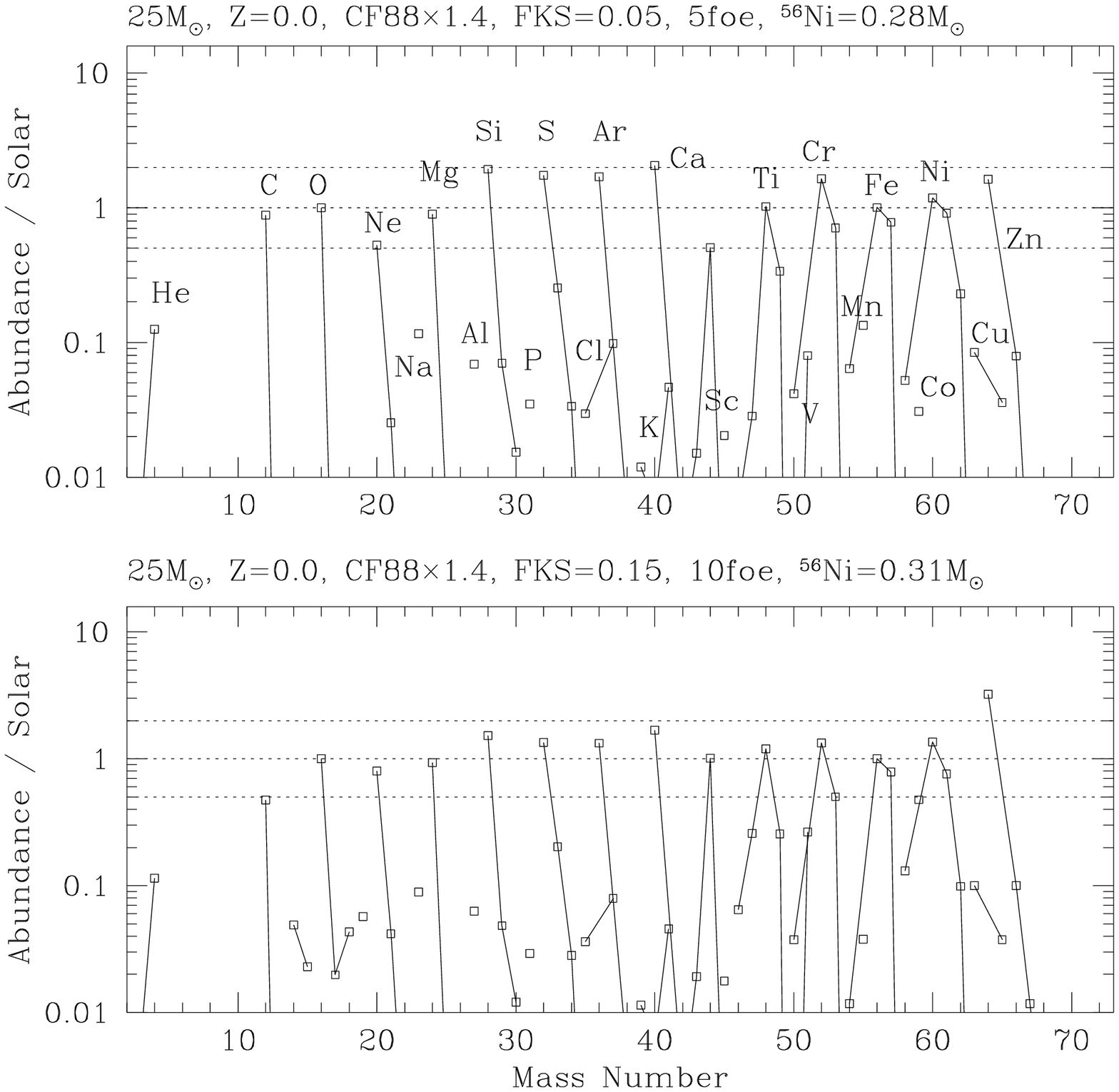}

\hskip 3cm
\epsfxsize=12.cm
\epsfysize=11.cm
\epsfbox{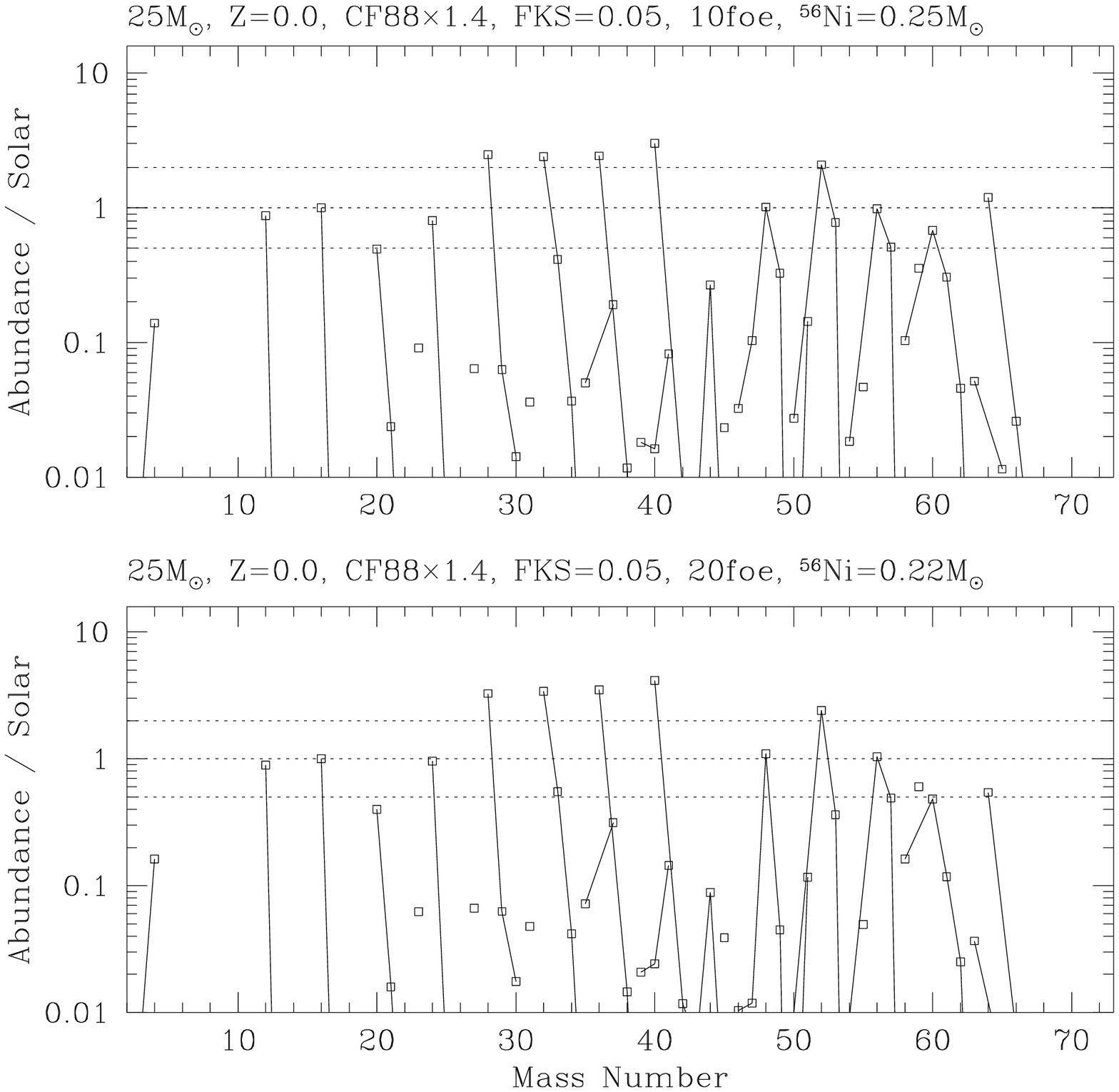}

\caption{Parameter dependence of the 25 $M_\odot$, $Z=0$ model yields.
 From top to the bottom, explosion energy is varied as 5, 10, 10 and 20
foe (= 10$^{51}$ erg). Mass cut is chosen for [$^{16}$O/$^{56}$Ni]
= 0. The differences of the 2nd and 3rd panels are in the choice of
convective mixing parameter $f_k$.  Second one is a case for more
efficient mixing.
\label{abunparam}}
\end{figure}

  Figures \ref{abunz002}-\ref{abunparam} show the integrated
abundances of stable isotopes in the ejecta relative to the solar
values (Anders \& Grevesse 1989) for the stars of $M = 13-25$ \ms and
$Z=0$ and 0.02. (Here we adopt the lower efficiency of mixing,
$f_k=0.05$.)  Each yield is normalized to the solar $^{16}$O value.

\begin{figure}
\hskip 3cm
\epsfxsize=12.cm
\epsfysize=11.cm
\epsfbox{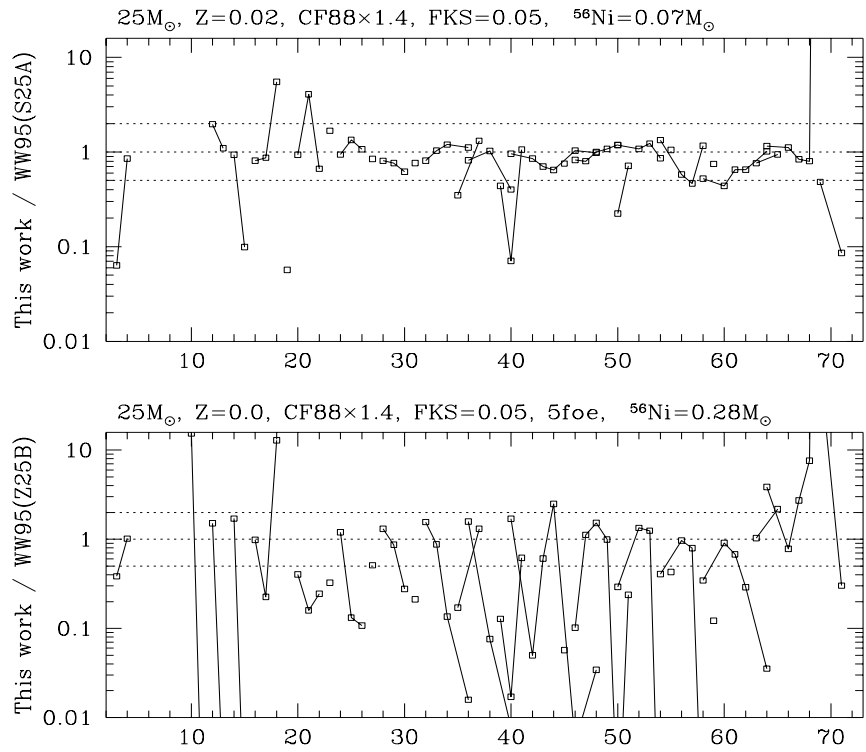}

\caption{Comparison of this work with Woosley \& Weaver (1995).
\label{abuncomp}}

\end{figure}

 Figure \ref{abunz002} shows the yields from the solar metallicity
stars. The explosion energy, i.e., the final kinetic energy, is
assumed to be $E= 10^{51}$ erg and the mass cut is chosen to eject
0.07 \ms $^{56}$Ni.  The abundance ratios are mostly consistent with
the solar ratios from C to Ca, but show significant deviations from
the solar ratios for heavier species.  Further parameter study is
necessary to see how the results depend on the metallicity, $f_k$, and
the mass cut.  The upper panel of Figure \ref{abuncomp} shows that our
result for $f_k=0.05$ is roughly consistent with WW95, though our
$^{12}$C$(\alpha,\gamma)^{16}$O rate is smaller than WW95 by a factor
of 1.2.

 The results for the metal-free models are shown in Figure
\ref{abunz0}. A distinctive feature of the $Z=0$ models compared with
the $Z=0.02$ models is that, alpha nuclei (from C to Zn) are much more
abundant than others. In Figure \ref{z0z002comp} we show the yield
ratios between $Z=0$ and $Z=0.02$.  The most evident signature of the
Pop III abundance pattern appears in the large isotopic ratios for the
even $Z$ elements (such as $^{13}$C/$^{12}$C) and the deficiency of
the odd $Z$ elements such as $^{23}$Na, $^{27}$Al, and $^{31}$P. We
will come back to this point in Section \ref{pop3abun}. Comparison of
our results with WW95 in the lower panel of Figure \ref{abuncomp}
shows that our model has smaller abundances of the even $Z$
neutron-rich isotopes; this is probably because WW95 included neutrino
irradiation effects which we do not.
 
 Isotopes heavier than Si are mostly produced during explosive
nucleosynthesis. The explosive nucleosynthesis depends on several
parameters such as the explosion energy and mass cuts which are not
well determined because of the uncertainty in the explosion mechanism.
Some of the parameter dependences for $Z=0$ are shown in Figure
\ref{abunparam}.

 The difference in the bottom panel of Figure \ref{abunz0} and the top
panel of Figure \ref{abunparam} is the mass cut. In the latter, deeper
mass cut is chosen so that [$^{16}$O/$^{56}$Ni] = 0; then
interestingly the abundance ratios for the even $Z$ elements up to Zn
can be roughly solar. Elements heavier than Zn might also be produced
in the same way though we have not calculated. 

 The top, 2nd and 3rd panels of Figure \ref{abunparam} show the
dependence on the explosion energy. The influence of the higher
explosion energy is mainly to shift the distribution of the peak
temperature outward.  As a result, the location of the mass cut
affects drastically the produced amount of $^{56}$Ni. These figures
show that Si and Ca are more abundant for a larger explosion energy.

 Finally the effect of more efficient convective mixing is shown in the
2nd panel of Figure \ref{abunparam}. It appears that this parameter
choice ($f_k=0.15$) is better to reproduce the solar abundance ratio for
the even $Z$ elements. However, we need further calculations for other
masses and parameters in order to find general trends.

\section{Cr, Mn, Co, and Zn}

 As mentioned in \S1, recent high resolution abundance surveys
discovered interesting trends of [Cr/Fe], [Mn/Fe] and [Co/Fe] with
[Fe/H], that is, both [Cr/Fe] and [Mn/Fe] decreases as metallicity
declines for [Fe/H] = $-2.4$ to $-4.0$ while [Co/Fe] increases
(McWilliam \etal 1995ab; Ryan \etal 1996; McWilliam 1997).  One might
suspect that these trends would be a signature of Pop III.

\begin{figure}
\hskip 3cm
\epsfxsize=12.cm
\epsfysize=11.cm
\epsfbox{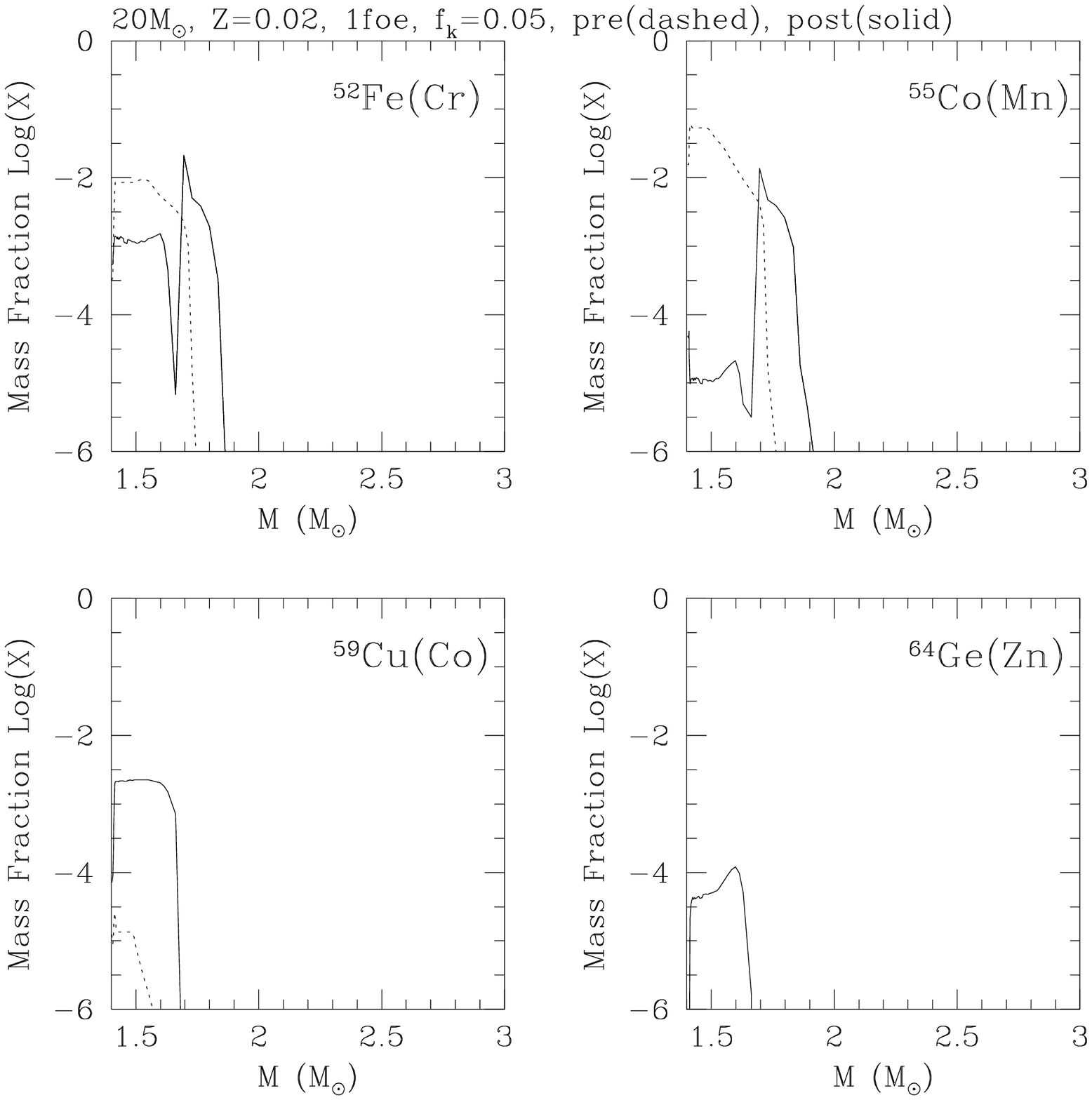}

\hskip 3cm
\epsfxsize=12.cm
\epsfysize=11.cm
\epsfbox{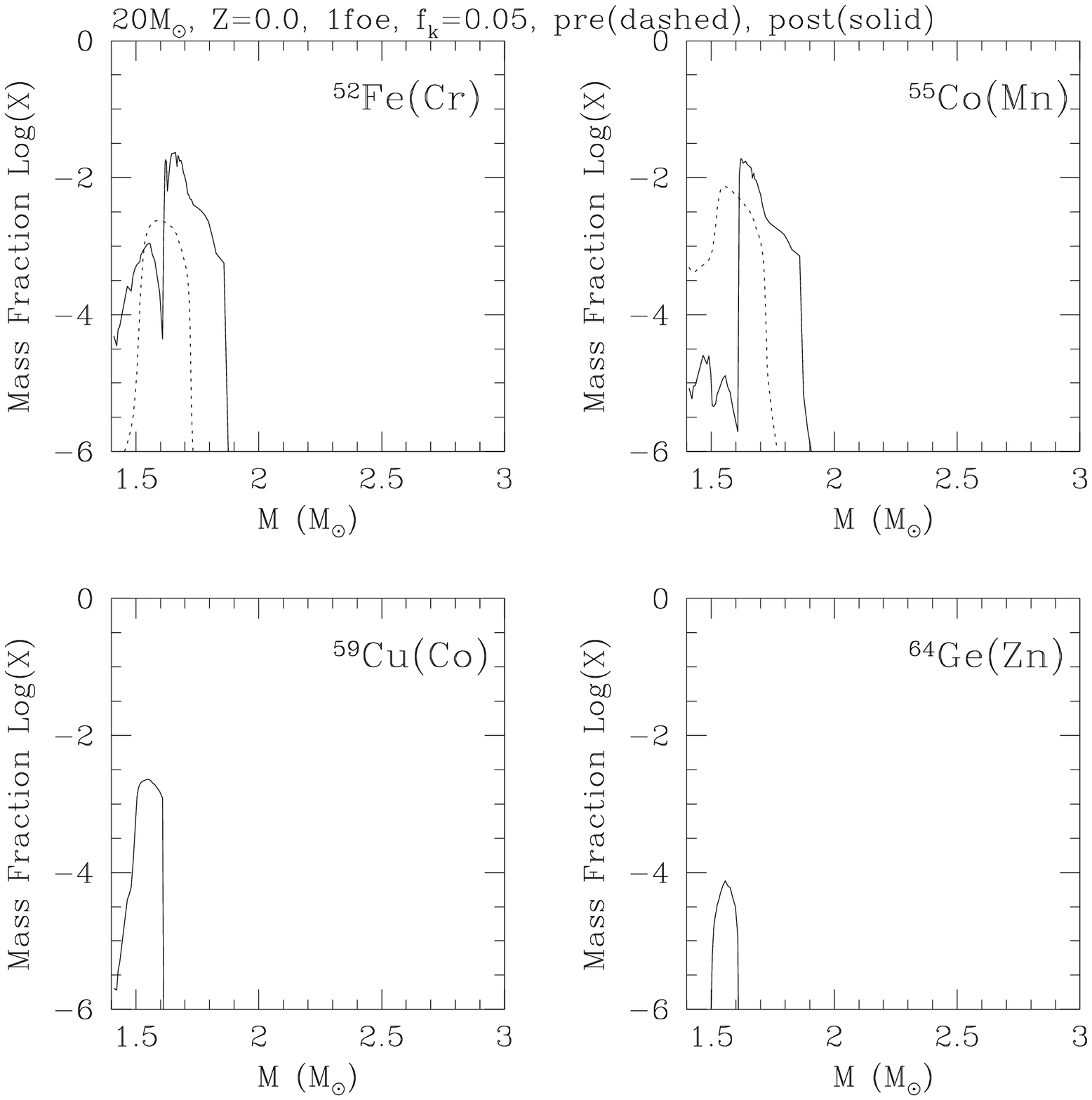}
\caption{Pre and post-explosive abundance distribution of $^{52}$Fe,
$^{55}$Co, $^{59}$Cu, $^{64}$Ge for the 20 \msp, $Z=0.02$ (upper 4
panels) and $Z=0$ (lower 4 panels) models.  These unstable isotopes
are main sources to decay into Cr, Mn, Co and Zn, respectively.
\label{dist4}}
\end{figure}

 These trends could be explained with SNe II, if these abundance
ratios strongly depend on metallicity.  Figure \ref{dist4} shows that
the amount of iron peak elements produced by the explosive synthesis
such as Cr, Mn, Co do not much depend on the initial stellar
metallicity (even though Mn and Co have odd $Z$). Although $Y_e$ for
$Z=0$ and 0.02 models is slightly different, these results show that
the effect is not very large.

 Therefore, other factors would be responsible to cause the above
trends.  In order to investigate the ratios [Cr/Fe], [Mn/Fe] and
[Co/Fe] we look into the regions where these elements are produced.
First of all, $^{56}$Ni, which decays into the most abundant Fe
isotope $^{56}$Fe, is produced not only in the complete Si-burning
region but also in the incomplete Si-burning region.  $^{59}$Cu, which
decays into $^{59}$Co, is produced in the complete Si-burning region.

$^{52}$Fe and $^{55}$Co, which decay into $^{52}$Cr and $^{55}$Mn,
respectively, are mostly synthesized in the incomplete Si-burning
region.  $^{52}$Fe is also synthesized in the complete Si-burning
region, but not a large amount.  Note that $^{55}$Mn and $^{59}$Co are
the only stable isotopes of these elements and therefore, their
abundances are identical with the Mn and Co element abundances.

The above discussion suggests that the choice of the mass cut can
affect the ratios [Cr/Fe], [Mn/Fe], and [Co/Fe]. For a deeper mass
cut, the ejected mass of the complete Si-burning region is larger
(i.e., the masses of Fe and Co are larger), while the ejected mass of
the incomplete Si-burning region remains the same (i.e., the masses of
Cr, Mn, and Fe are the same).  Accordingly the ratios of [Cr/Fe] and
[Mn/Fe] are smaller and [Co/Fe] is larger.  For a mass cut at larger
radii, these ratios show the opposite tendency.  Therefore, specific
choices of mass cuts in SNe II might explain the behavior of [Cr/Fe],
[Mn/Fe], and [Co/Fe] in the metallicity range $-4\le$ [Fe/H] $\le-2.5$
(Nakamura \etal 1999).

 The dependence of the abundance ratios on the mass cut is seen from
the comparison between Figures \ref{abunz0} and \ref{abunparam}.  The
relative thickness between the complete and the incomplete Si burning
layers may also depend on the efficiency of convective mixing $f_k$.

 Note also that Zn is produced from complete Si burning as seen in
 Figures \ref{dist4}.  In order to reproduce [Zn/Fe] $\sim$ 0 as
observed in metal-poor stars, the ejection of relatively large amount
of complete Si burning material is necessary as seen in Figure
\ref{abunparam} and discussed in \S6.  This would occur if the mass
cut is as deep as discussed above (Nakamura \etal 1999) or if such
complete Si burning material is preferentially ejected in a form of a
jet or bullets (Umeda 1999) from the deepest layer.

\section{Pop III characters in the abundance pattern}

\label{pop3abun}

 Massive Pop III stars no longer exist in the present universe.
However, if Pop III supernovae induced the formation of low-mass stars,
such stars should still alive and may be observable. Indeed some of
the observed metal-poor halo stars may be such stars. If we can
identify such stars, they are very useful in understanding the
nucleosynthesis in the earliest universe. Also we may even be able to
determine the typical mass range of Pop III stars by comparing theory
with observed abundance pattern.

 In the previous section we saw that the most distinctive features of
the Pop III nucleosynthesis compared with Pop I and II are the
abundance ratios in even $Z$ isotopes (e.g., $^{12}$C/$^{13}$C) and
the deficiency of the odd $Z$ elements such as $^{23}$Na, $^{27}$Al,
and $^{31}$P compared with even $Z$ elements.  In this section, we
discuss in more detail the metallicity dependence of the abundance
pattern to find specific characteristics of Pop III abundance pattern. 
As described above, yields from explosive nucleosynthesis depends on
the non-well-understood properties of explosion such as the mass cut,
mixing, etc. Therefore, we concentrate on the elements which are not
sensitive to the explosive synthesis.

\subsection{CNO elements}

\begin{figure}
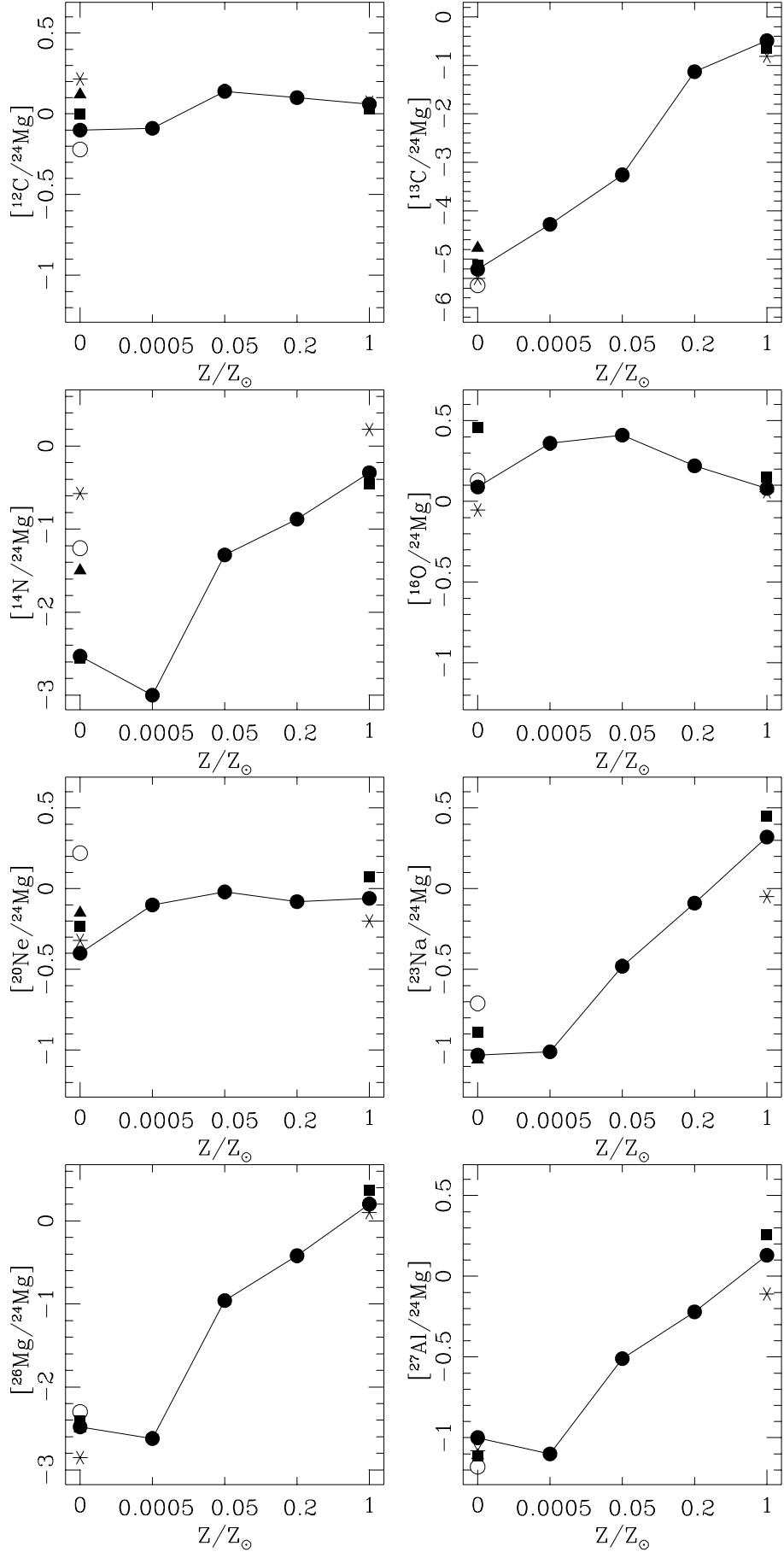

\hskip 3cm
\epsfxsize=12cm
\epsfysize=6cm
\epsfbox{umenomofig14a.epsi}

\hskip 3cm
\epsfxsize=12cm
\epsfysize=6cm
\epsfbox{umenomofig14b.epsi}

\hskip 3cm
\epsfxsize=12cm
\epsfysize=6cm
\epsfbox{umenomofig14c.epsi}

\hskip 3cm
\epsfxsize=12cm
\epsfysize=6cm
\epsfbox{umenomofig14d.epsi}

\caption{Metallicity dependence of the integrated abundance ratio 
of elements X to $^{24}$Mg relative to the solar ratio.
\label{abunrat}}

\end{figure}

 In order to study the metallicity dependence of the yields in more
detail, we calculate stellar evolution and explosive nucleosynthesis
for the 20 \ms models with $Z=0.02$, 0.004, 0.001, $10^{-5}$ and
0. The integrated abundance ratio of some elements X to $^{24}$Mg
relative to the solar ratio are shown in Figures \ref{abunrat}, where
[X/Y]$\equiv \log_{10}(X/Y)-\log_{10}(X/Y)_\odot$.  Here $^{24}$Mg is
used for normalization, because yield of $^{24}$Mg is independent of
mass cut and does not significantly depend on metallicity (e.g.,
0.066, 0.054, 0.040, and 0.081 \ms for $Z=0.02$, 0.004, 0.001 and 0,
respectively for the 20 \ms models); furthermore observationally
[$^{24}$Mg/H] appears to roughly correlate with metallicity [Fe/H]
(Shigeyama \& Tsujimoto 1998).  In these figures, the filled circles
show the ratios for the 20 \ms star. The asterisk, triangle, and
square represent 13, 15, 25 \ms models, respectively (only for $Z$=0
and 0.02), and open circle is the 25 \ms $Z=0$ model with more
efficient convective mixing ($f_k=0.15$).

 These figures show that the abundance ratios for $^{12}$C and
$^{16}$O are almost independent of metallicity ($Z$), while the ratios
for $^{13}$C and $^{14}$N are smaller for smaller metallicity. 

 The result for $^{14}$N yield is interesting.  The
[$^{14}$N/$^{24}$Mg] ratio decreases as metallicity gets smaller, but
its metallicity dependence is sensitive to the stellar mass.  For
$Z=0$, [$^{14}$N/$^{24}$Mg] is larger for less massive stars and its
value for the 13 \ms model is only 1/4 of the solar ratio.  Therefore,
primordial $^{14}$N production from Type II supernovae may be
significant.  This $^{14}$N production mechanism by these stars are
similar to the suggestion in WW95. There, they commented that in some
of their unpublished early calculations the convective He-shell
penetrated into the H-layer with the consequent production of large
amounts of primary $^{14}$N. In our model also $^{14}$N is produced in
the He-burning shell during central C-burning. In this stage, the
He-shell burning becomes strong enough to form a convective layer.
This convective layer reaches the H-rich layer and the CNO-cycle
operates to produce $^{14}$N. In our model this mechanism is specific
to the Pop III stars.  For larger metallicity, the convective layer in
the He-burning shell is hard to reach the H-rich region without
overshooting because H-shell burning is so active to form a high
entropy barrier.  The $^{14}$N production may be enhanced if we assume
more efficient convective mixing as shown by the example in Figure
\ref{abunrat} (open circle).

\subsection{Odd $Z$ elements}

 The abundances of odd $Z$ elements produced during C-burning such as
$^{23}$Na, $^{27}$Al, and $^{31}$P may be good indicators for the
metallicity of the progenitor stars, because these elements are
secondary and their abundances are expected to be smaller for smaller
metallicity.  Figure \ref{abunrat} shows that $^{23}$Na and $^{27}$Al
abundances are smaller for smaller metallicity, in contrast to the
primary elements such as $^{20}$Ne and $^{24}$Mg. These figures show,
however, that the declines of [$^{23}$Na/$^{24}$Mg] and
[$^{27}$Al/$^{24}$Mg] toward smaller metallicity almost saturate for
$Z/Z_\odot \lsim 5\times 10^{-4}$. Hence it might be difficult to
distinguish the Pop III abundance pattern from the abundance pattern
of very low-metal Pop II stars only by these elements.  Nevertheless
we may say that the stars with the lowest [$^{23}$Na/$^{24}$Mg] and
[$^{27}$Al/$^{24}$Mg] are most likely composed of the lowest
metallicity SN ejecta.

\begin{figure}
\hskip 3cm
\epsfxsize=12cm
\epsfysize=10cm
\epsfbox{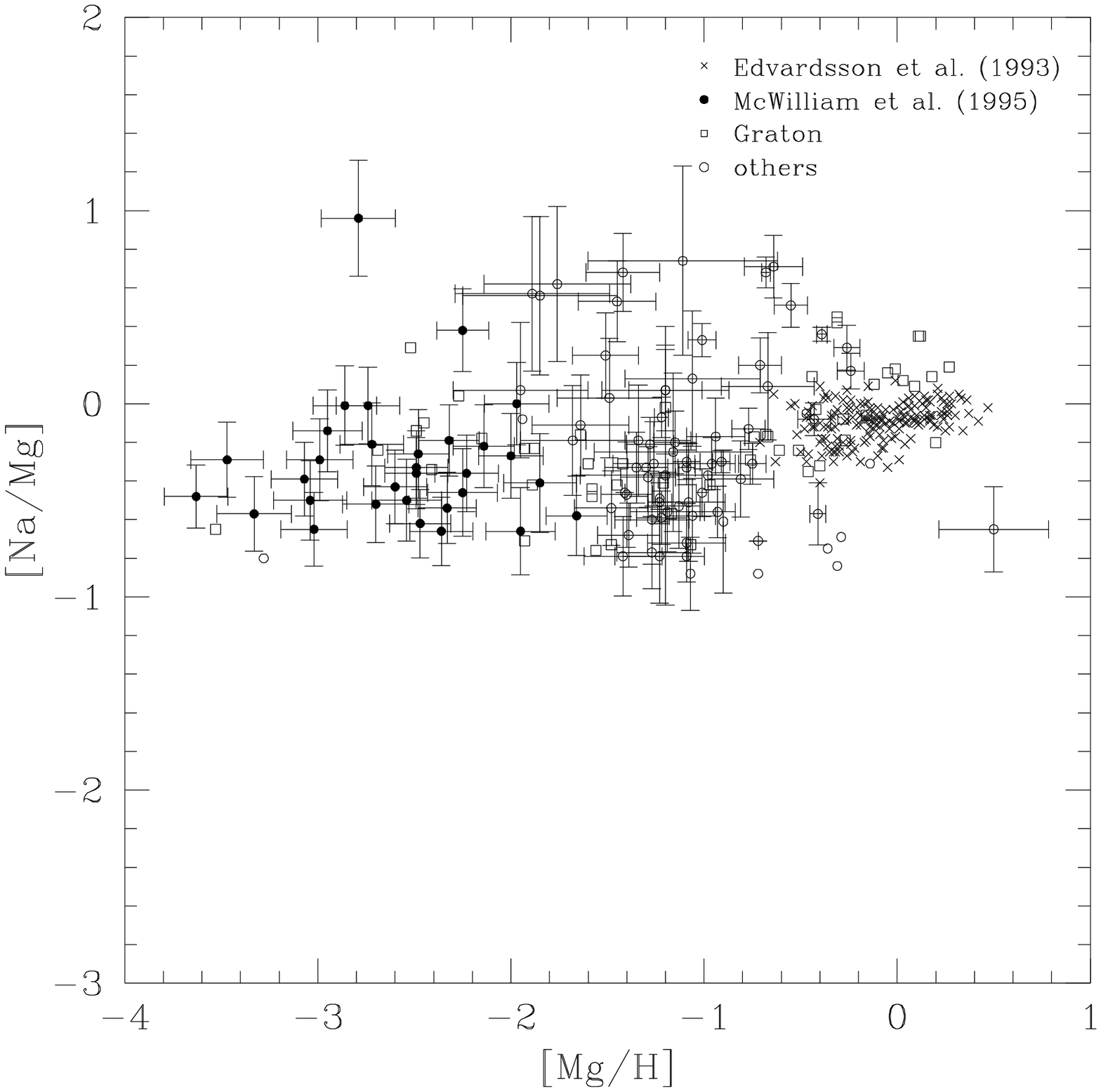}

\hskip 3cm
\epsfxsize=12cm
\epsfysize=10cm
\epsfbox{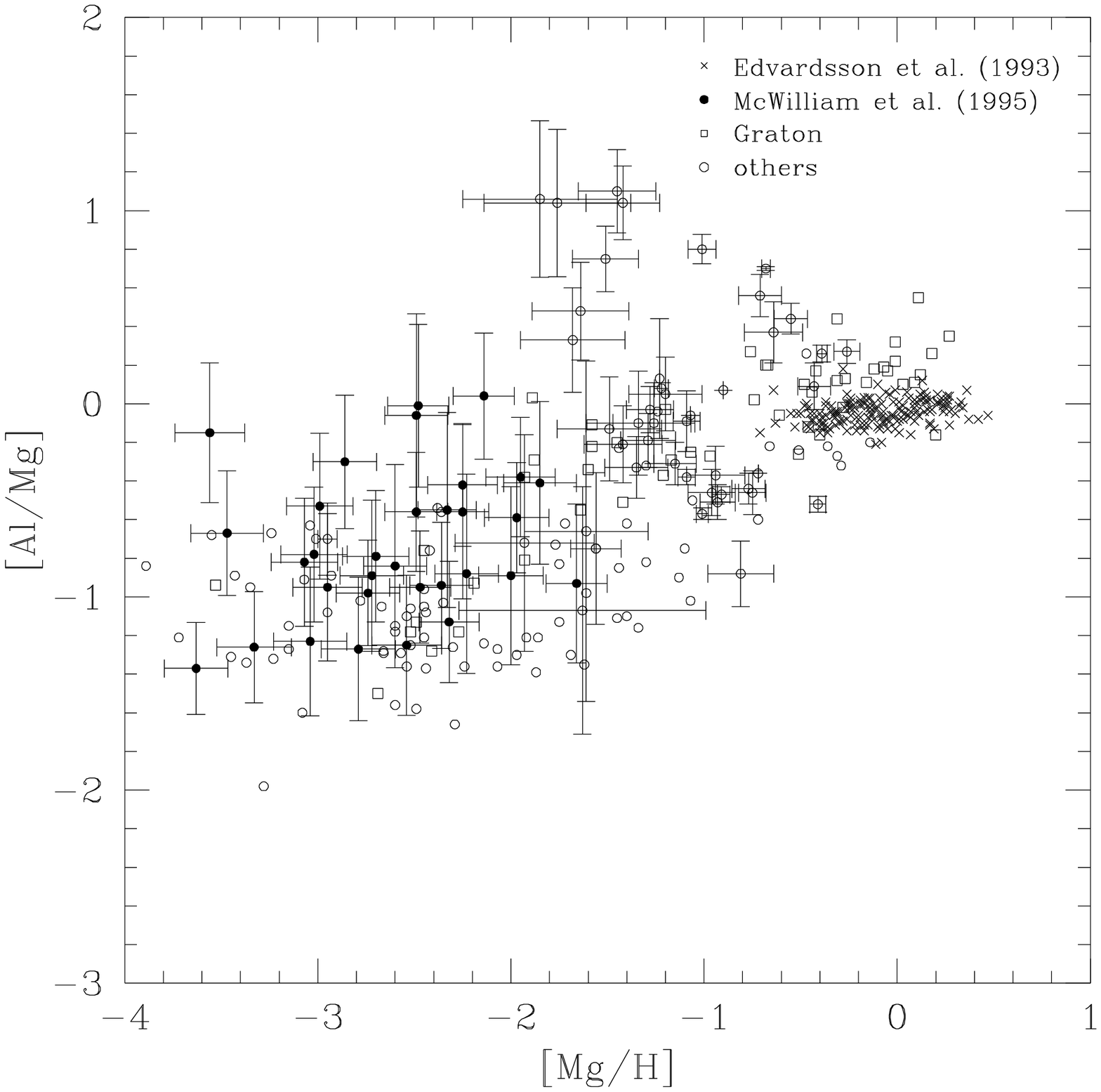}

\caption{Observed abundance ratios [Na/Mg]
and [Al/Mg] in stars of the halo and the local disk.
\label{abunobs}}

\end{figure}

 Figure \ref{abunobs} shows the observed abundance ratios [Na/Mg] and
[Al/Mg] in stars of the halo and the local disk (McWilliam \etal
1995ab; Ryan \& Norris 1991; Edvardson \etal 1993). The stars with the
Pop III abundance pattern, if they exist, should belong to the
low-metal stars ([Mg/H] $\lsim - 2$), but not necessarily the lowest
metallicity stars because of the following reason.  Such stars are
composed of Pop III SNe ejecta mixed with interstellar matter (ISM).
The amount of ISM mixed with the SN ejecta is estimated by Ryan \etal
(1996). It depends on several uncertain factors but typically the
metallicity of the mixed ejecta ranges from [Mg/H]$\sim -4$ to $-2$
(see also Nakamura \etal 1999, Shigeyama \& Tsujimoto 1998).

 The observed patterns of [Na/Mg] and [Al/Mg] vs [Mg/H] are similar.
For [Mg/H] $\lsim -2$, the lower bounds are almost constant ($\sim -
1$) and upper bounds increase with [Mg/H].  The existence of the
constant lower bounds is consistent with our results shown in Figures
\ref{abunrat}; it can be explained if these stars are composed of
ejecta of Pop III or very low metal Pop II supernovae. Although only
 from these data it is hard to find the differences between the Pop III
and the very low metal Pop II abundance, or the differences of
progenitor masses, further theoretical and observational study may
make discrimination possible.

\subsection{Alpha Elements/Fe}

 In \S6 and Figure \ref{abunparam}, we discuss the cases with
[$^{16}$O/$^{56}$Ni] = 0.  These SNe produce a large amount of
$^{56}$Ni ($\sim 0.3 M_\odot$), which is comparable to hypernovae
SN1998bw and SN1997ef (Iwamoto \etal 1998, 1999) and even Type Ia
supernovae (Nomoto \etal 1984).

In this connection, the abundance pattern of the very metal-poor
binary CS22873-139 ([Fe/H] $= -$3.4) is interesting.  This binary has
only the upper limit to [Sr/Fe] $< - 1.5$, thus being suggested to be
a second generation star (Nordstrom \etal 1999).  Although [Al/Fe]
$\sim -0.6$ is too large for Pop III, other abundance features should
be examined.

Interesting pattern is that this binary shows almost the solar Mg/Fe
and Ca/Fe ratios.  This would not mean that Type Ia supernovae
contributed to the enrichment of Fe for such a metal-poor star.  
We rather suggest that the abundance pattern of this binary 
originates from hypernovae.  Another feature of CS22873-139 is
enhanced Ti/Fe
([Ti/Fe] $\sim +0.6$; Nordstrom \etal 1999), which also seems to be
the result from hyper-energetic or hyper-asymmetric explosion.

Since Pop III stars may consist of many massive stars that undergo
hypernova-like explosions, [alpha element/Fe] $\sim$ 0 could possibly
be a signature of Pop III SNe.

\section{Summary}

 We calculate pre-supernova evolutions and supernova explosions of
massive stars ($M=13-25 M_\odot$) for various metallicities.  We
find some characteristic abundance patterns of nucleosynthesis in
the metal-free (Pop III) stars. 

 The most distinctive features of the Pop III nucleosynthesis compared
with Pop I and II are that, for $Z=0$, alpha nuclei (from C to Zn) are
much more abundant than others. (The abundance pattern of
$\alpha$-nuclei can be similar to the solar abundance.) Therefore, by
looking for the smallest isotopic ratios for the even $Z$ elements
(such as $^{13}$C/$^{12}$C) or the smallest abundance of odd $Z$
elements in the metal-poor star abundances, we can find candidates of
the stars having Pop III abundance pattern.  Also, near solar ratios
of alpha elements/Fe might be a signature of Pop III which could
produce a large amount of Fe in hyper-energetic explosion.

The abundance ratios of odd $Z$ to even $Z$ elements such as Na/Mg and
Al/Mg become smaller for lower metallicity. However these ratios
almost saturate below $Z \lsim 10^{-5}$, and [Na, Al/Mg] $\sim -1$ for
Pop III and low metal Pop II nucleosynthesis. This result is
consistent with abundance pattern of metal poor stars, in which these
ratios also saturate around $-1$. We suggest that these stars with the
lowest [Na/Mg] or [Al/Mg] may contain the abundance pattern of Pop III
nucleosynthesis. 

Metal poor stars show interesting trends in the ratios of [Cr, Mn,
Co/Fe]. We discuss that these trends are not explained by the
differences in metallicity, but by the relative thickness between the
complete and the incomplete Si burning layers. Large [Co/Fe] and small
[Cr, Mn/Fe] values found in the observations are explained if mass cut
is deep or if matter is ejected from the complete Si burning layer in
a form of a jet or bullets.

We also find that primary $^{14}$N production occurs in the massive
Pop III stars, because these stars have radiative H-rich envelopes so
that the convective layer in the He-shell burning region can reach the
H-rich region.  Then primary $^{14}$N production by NO-cycle can
occur without overshooting.

Further study may make it possible to determine the metallicity and
progenitor mass ranges of a SN which produced the metal-poor stars.

\bigskip
 We would like to thank M. Hashimoto and H. Saio for providing us with
the basis of the present evolutionary code.  This work has been
supported in part by the Grant-in-Aid for by COE Scientific research
(07CE2002, 0980203) of the Japanese Ministry of Education, Science,
and Culture.



\end{document}